\documentclass[preprint,review,12pt]{elsarticle}

\def\Rone{} 

\usepackage{amssymb}
\usepackage{framed,multirow}
\usepackage{amsmath}
\usepackage{longtable}
\usepackage{makecell}
\usepackage{setspace}
\usepackage{hyperref}
\usepackage{pdflscape}

\journal{Expert Systems with Applications}

\begin{document}

\begin{frontmatter}



\title{Now and Future of Artificial Intelligence-based Signet Ring Cell Diagnosis: A Survey}


\author[a]{Zhu Meng\fnref{fn1}}\ead{bamboo@bupt.edu.cn}\author[a]{Junhao Dong\fnref{fn1}}\ead{djh1999@bupt.edu.cn}\author[b]{Limei Guo\fnref{fn1}}\ead{guolimei@bjmu.edu.cn}\author[a]{Fei Su}\ead{sufei@bupt.edu.cn}\author[a]{Jiaxuan Liu}\ead{liujiaxuan@bupt.edu.cn}\author[b]{Guangxi Wang}\ead{guangxiwang@bjmu.edu.cn}\author[a]{Zhicheng Zhao\corref{cor1}}\ead{zhaozc@bupt.edu.cn}

\affiliation[a]{organization={School of Artificial Intelligence},
            addressline={Beijing University of Posts and Telecommunications}, 
            city={Beijing},
            postcode={100876}, 
            country={China}}
\affiliation[b]{organization={Department of Pathology, School of Basic Medical Sciences},
            addressline={Peking University Third Hospital}, 
            city={Beijing},
            postcode={100191}, 
            country={China}}

\cortext[cor1]{Corresponding author.}
\fntext[fn1]{These authors contributed equally to this work.}
\begin{abstract}

Signet ring cells (SRCs), associated with a high propensity for peripheral metastasis and poor prognosis, critically influence surgical decision-making and outcome prediction. However, their detection remains challenging even for experienced pathologists. While artificial intelligence (AI)-based automated SRC diagnosis has gained increasing attention for its potential to enhance diagnostic efficiency and accuracy, existing methodologies lack systematic review. This gap impedes the assessment of disparities between algorithmic capabilities and clinical applicability. \Rone{This paper presents a comprehensive survey of AI-driven SRC analysis from 2008 through June 2025. We systematically summarize the biological characteristics of SRCs and challenges in their automated identification. Representative algorithms are analyzed and categorized as unimodal or multi-modal approaches. Unimodal algorithms, encompassing image, omics, and text data, are reviewed; image-based ones are further subdivided into classification, detection, segmentation, and foundation model tasks. Multi-modal algorithms integrate two or more data modalities (images, omics, and text).} Finally, by evaluating current methodological performance against clinical assistance requirements, we discuss unresolved challenges and future research directions in SRC analysis. This survey aims to assist researchers, particularly those without medical backgrounds, in understanding the landscape of SRC analysis and the prospects for intelligent diagnosis, thereby accelerating the translation of computational algorithms into clinical practice.
\end{abstract}


\setcounter{page}{1}


\end{frontmatter}


\section{Introduction}
\label{sec:introduction}
Signet ring cells (SRCs) reflect special histopathological features where nuclei are squeezed into eccentrically placed crescent shapes by abundant intracellular mucins \citep{S0}. Histopathologically, SRC carcinoma is noted when more than 50\% tumor cells contain SRC features, while tumors with few SRC components are also concerned \citep{S0}. Although the majority of SRC carcinomas occur in the gastrointestinal tract, they also appear in esophagus, lung, pancreas, appendix, gallbladder, breast, urinary bladder, ovary, prostate, skin, and other tissues \citep{S1,S2,S3,S11,S12}. When gastric SRC carcinoma is diagnosed, it is aggressive and can be accompanied by diffuse growth of adenocarcinoma cells and extensive connective tissue proliferative response \citep{S10}, especially when infiltrating into the layers of submucosa, muscularis propria or serosa. SRC features are associated with high peripheral metastasis rate, poor response to neoadjuvant treatment, and particular dismal survival \citep{S1,S2,S4,S5,S6,S7,S8}. SRCs will not aggregate to form a relatively regular structure, and thus, they are difficult to identify in low magnification pathological diagnosis, while the cell morphology in high magnification pathological images is very similar to plasma cells, intestinal metaplasia, and activated endothelial cell \citep{AS2}. Therefore, SRCs are easily missed even for experienced pathologists. As a result, computer-aided algorithms are expected to assist pathologists to improve the screening speed and accuracy of SRCs.

\Rone{Computational approaches leveraging omics data, including genomics, transcriptomics, and proteomics, have historically favored traditional machine learning algorithms for SRC profiling. These algorithms like random forests gained prominence due to their computational tractability and relative explainability in identifying molecular signatures associated with SRC aggressiveness and metastasis} \citep{O-Pro-RF,O-biology,O-pancr}. \Rone{Recent advances, however, witness a paradigm shift: deep neural networks now model complex omics interactions beyond machine learning’s capability} \citep{O-DOMSCNet}, \Rone{while emerging foundation models (e.g., multi-omic pre-trained transformers)} \citep{T-Omics-LLM} \Rone{enable cross-modal knowledge transfer for predicting SRC phenotypes.} 

Since artificial intelligence (AI) has made remarkable achievements in natural image processing such as classification, segmentation, and detection \citep{R-classification, R-detection, R-Segmentation}, more and more clinically effective medical image processing algorithms based on convolutional neural networks (CNNs) have emerged \citep{AF6, Google-LYNA, S13, S14}. Actually, the automatic diagnosis of SRCs has been concerned since 2008, where a deep learning network, LeNet \citep{LeNet}, was combined with color features to detect SRCs \citep{A2008}. However, the automatic screening algorithms have not made a greater breakthrough until recent years. Different from normal cells and other tumor cells, the distributions of SRC are various, which leads to the difficulty for the algorithms to capture typical features. And thus, SRC-related data were often ignored or removed in some deep learning-based lesion screening tasks. For example, Cowan et al. excluded slides suggestive of SRC carcinoma because these entities performed poorly in their training data \citep{R1}. Although the Lizard Dataset contained nearly half a million labeled nuclei in hematoxylin and eosin (H\&E) stained colon tissue, SRCs were not involved and expressed particular interest in future work \citep{R3}. Actually, SRCs have gradually received attention in recent years. SRC carcinoma was often mixed in the automatic identification of whole-slide images (WSIs) as one of the subtypes, and was often analyzed as typical cases \citep{R7, R9, AF5, B2364}. However, many studied suggested that the sensitivity of identifying SRC lesions was significantly lower than that of well-differentiated adenocarcinoma without SRCs \citep{R6, AF2, AF3, AF6, AF8, B2328}. Therefore, more attention needs to be paid to the recognition of SRCs, which was indeed concerned in the discussions of some articles \citep{AS1, AL1, AL3, B20303, B202326, B2335, B202377}. Particularly, the DigestPath dataset \citep{DigestPath} was released to encourage typical SRCs detection in histopathology image patches.

\Rone{Beyond unimodal approaches confined to either histopathology images or omics data, pioneering studies have responded to AI's evolving landscape by developing integrated multi-modal frameworks. These collaborative methodologies primarily fuse complementary data streams, notably histopathology images with clinical text} \citep{M2-CHIEF,M2-MUSK,M2-Prov-GigaPath}, \Rone{images with omics profiles} \citep{M1-MISO,M1-OmiCLIP}, \Rone{or tripartite integrations of images, text, and omics} \citep{M3-GECKO,M3-ModalTune,M3-PCA,M3-spEMO}, \Rone{often leveraging the potent representational capacity of large language models (LLMs). Such multi-modal architectures align fundamentally with contemporary clinical diagnostic paradigms that mandate multi-factorial analysis, representing a conceptual advance in intelligent systems for SRC characterization and diagnosis.}

\begin{figure*}[!t]
	\centering
	\includegraphics[width=0.8\textwidth]{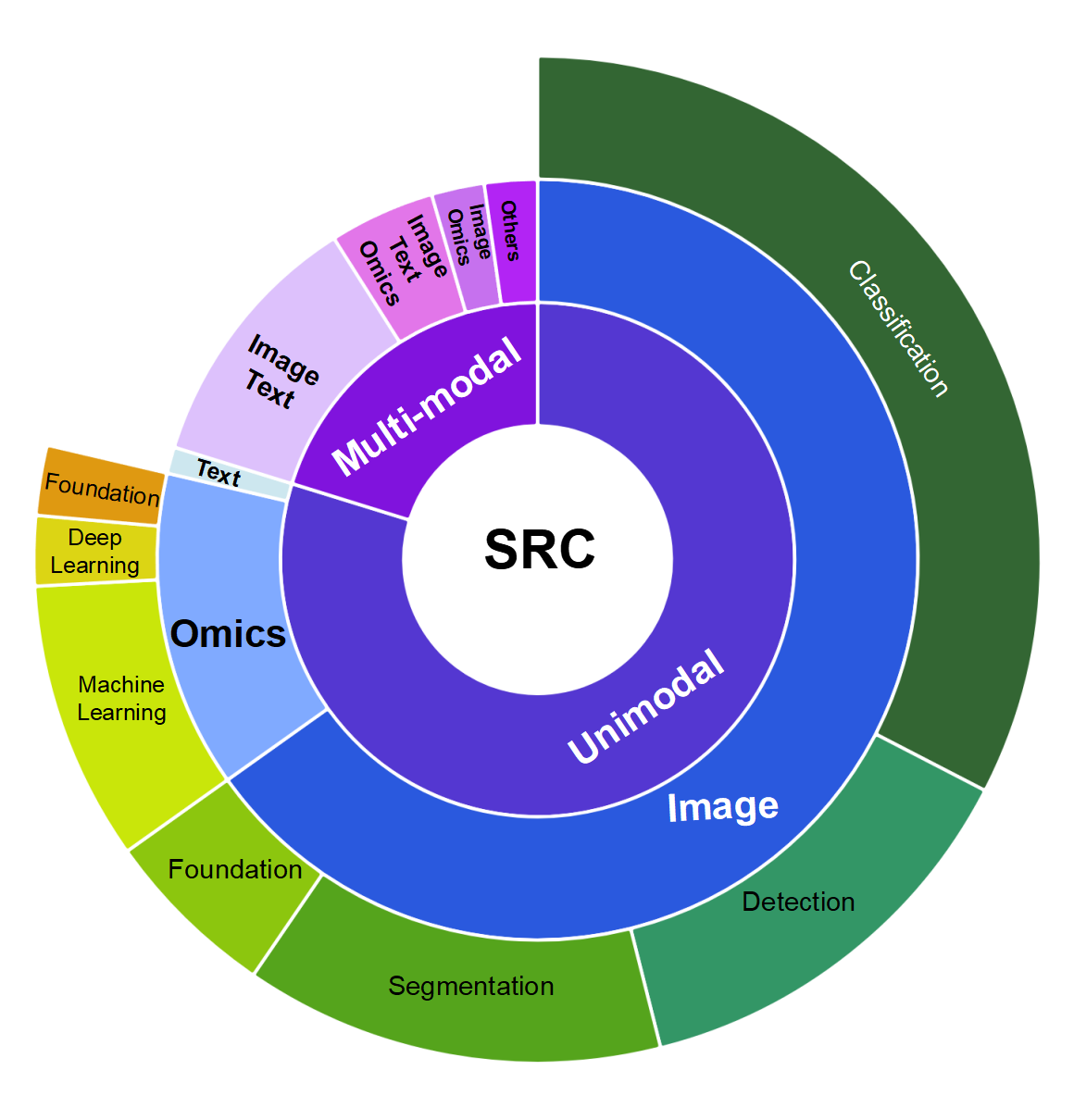}
	\caption{\label{fig:ALL}\Rone{An overview of the SRC diagnosis studies driven by AI.}}
\end{figure*}

We observe that there has been a representative and wide-ranging survey for AI-based medical image processing \citep{ZS5}, especially for microscopy and histopathology images \citep{ZS8, ZS9, ZS2, ZS3, ZS4, ZS26}. In addition, surveys of AI algorithms for identification for different tissues and organs were also emerged, including brain \citep{ZS28}, lung \citep{ZS7}, cervix \citep{ZS6}, colorectum \citep{ZS1}, skin \citep{ZS21}, and liver \citep{ZS25}. However, SRCs were overlooked in these reviews. Therefore, this paper summaries most of the AI-related articles for SRC identification until \Rone{June 2025} to help researchers comprehensively understand this field. The distribution of the included studies is illustrated in Fig. \ref{fig:ALL}.

The remainder of this paper is organized as follows. Section \ref{sec:overview} presents the overview of problem description, public SRC datasets (the corresponding evaluation metrics are summarized in the Supplementary Materials), and the challenges of automatic diagnosis for SRCs. Section \ref{sec:unimodal} \Rone{comprehensively analyzes single-modality methodologies, categorizing them into following principal domains: image pre-processing techniques, image-based classification/detection/segmentation approaches, histopathology foundation models, omics-driven computational frameworks, and clinical text mining methods.} Section \ref{sec:multimodal} \Rone{subsequently examines synergistic multi-modal integrations, specifically investigating four paradigm categories: image-text clinical report collaborations, histopathology-omics collaborations, tripartite image-text-omics collaborations, and other multi-modal combinations.} Section \ref{sec:discussion} discusses the limitations of existing algorithms and the future trends for clinically accurate SRC identification with AI assistance. Section \ref{sec:conclusions} provides the conclusions of this paper.

\section{Overview}
\label{sec:overview}
\subsection{Problem description}
\label{subsec:problem}

\Rone{Computational SRC diagnosis fundamentally constitutes learning a mapping function $f_\theta: \mathcal{X} \rightarrow \mathcal{Y}$, where $\mathcal{X} = ({\mathcal{X}_{\text{img}}, \mathcal{X}_{\text{omics}}, \mathcal{X}_{\text{text}}})$ denotes the input space spanning SRC-related medical images, omics profiles, and clinical text. For the unimodal mapping algorithms, among the three elements of the input $\mathcal{X}$, exactly one element is non-zero. $\mathcal{Y}$ represents diagnostic outputs, and  $f_\theta$ is the unimodal or multi-modal mapping algorithm with parameters $\theta$. 

Given this survey's detailed analysis of image-based algorithms, we specifically formalize problem description for the image processing algorithms.} Apart from a particular study on single-shot femtosecond stimulated Raman scattering \citep{B2327}, most SRC diagnosis algorithms were carried out based on four types of images, namely, H\&E stained WSIs, computed tomography (CT), magnetic resonance imaging (MRI), and endoscopic images. Since histopathological images at high magnification clearly show the morphology of nuclei and cytoplasm after staining, the automatic algorithms of SRCs based on H\&E stained WSIs are notably more than the other three types. Following the natural image analysis pipelines with high-performance, the SRC diagnosis algorithms can also be divided into three categories: classification, detection and segmentation.

\textbf{Image classification:} For classification, the goal is to train a model $f_\theta$ to predict the class label of an arbitrary test sample (e.g., a WSI or a patch). Specifically, in supervised learning, we define the training set as $Tr=\left(x_{tr}^{\left(i\right)},y_{tr}^{\left(i\right)}\right)_{i=1}^{N_{tr}}$, where $N_{tr}$ is the total number of training samples. For each sample, $x_{tr}^{\left(i\right)}\in R^{C_h\times H\times W}$ denotes the image to be classified with a resolution of $H\times W$ and $C_h$ channels, and $y_{tr}^{\left(i\right)}\in\ R^{C_l}\ $ denotes the ground-truth annotation with $C_l$ categories. Binary classification is the most common scenario for SRC diagnosis, where SRCs are usually regarded as positive samples. For instance, to predict whether a patch of WSIs contains tumors, the label is set to 1 for a positive case and 0 for a negative one. During the training process, a classification model $f_\theta$ is trained with the training set $Tr$ based on a specific loss function $L$. For inference process, the trained model provides the prediction $p_{te}^{\left(i\right)}$ for each sample $x_{te}^{\left(i\right)}$ in the test set. In addition, the parameters pretrained on large-scale datasets and slide-level annotations are also leveraged for other training strategies, such as transfer learning and weakly supervised learning. The details of SRC-related classification algorithms are elaborated in Section \ref{subsec:classification}.

\textbf{Image detection:} In object detection, the model $f_\theta$ is trained to perform both object localization and classification. Therefore, the given annotations are modified as $y_{tr}^{\left(i\right)}=\left(b_j^{\left(i\right)},c_j^{\left(i\right)}\right)_{j=1}^{M_i}$, where $b_j^{\left(i\right)}$ represents the position and size of the $j$th object’s bounding box in the $i$th image, while $c_j^{\left(i\right)}$ contributes its category, and $M_i$ represents the number of objects in this image. In SRC diagnosis, the aim is to train a detector $f_\theta$ with a classification loss $L_{cla}$ and a regression loss $L_{reg}$, which can accurately locate the SRCs in the test set. Besides, semi-supervised learning strategy (see Section \ref{subsec:detection}) is also applied to handle the incomplete labeling problem of the SRC detection.

\textbf{Image segmentation:} This task aims to achieve pixel-level dense prediction instead of simple classification of images. More formally, the ground-truth mask of training data is defined as $y_{tr}^{\left(i\right)}={(y_j^{\left(i\right)})}_{j=1}^{H\times W}$, where $y_j^{\left(i\right)}$ is the $j$th pixel-level annotation of the $i$th image, and the number of pixels is $H\times W$. For pathological images, the trained model $f_\theta$ is expected to generate reliable masks with densely predicted labels, on which sub-tasks can be performed such as survival prediction, lesion localization and cell segmentation. In addition to fully supervised learning, training strategies such as weakly supervised learning, collaborative learning and multi-task learning have also been introduced due to the lack of suitable dense annotations, which are illustrated in Section \ref{subsec:segmentation}.


\subsection{SRC datasets}
\label{subsec:SRC-datasets}
According to the articles covered in this survey, both public and private datasets were enrolled in SRC automatic diagnostic tasks. Among them, two public datasets DigestPath \citep{DigestPath} and The Cancer Genome Atlas (TCGA)\footnote{\url{https://portal.gdc.cancer.gov/} (Accessed July 21, 2025)} were most frequently adopted.
\subsubsection{DigestPath dataset}
DigestPath was a dataset from Digestive-System Pathological Detection and Segmentation Challenge 2019\footnote{\url{https://digestpath2019.grand-challenge.org/Dataset/} (Accessed July 21, 2025)}, which was the first public dataset for object detection of SRCs. It was applied to SRC automatic diagnosis of classification \citep{AL1, AL3}, detection \citep{AD1,AD2,AD3,B2321}, and segmentation \citep{AS1, AS2, B2346}. The training set was collected from two organs, i.e., intestinal and gastric mucosa, and was divided into positive and negative subsets. The positive one consisted of 77 pathological images cropped from 20 WSIs, which contained SRCs annotated with bounding boxes by experienced pathologists. The negative one consisted of 378 images from 79 WSIs with SRCs, but may contain other types of tumor cells. The sizes of the negative images were ($2000px\times2000px$), while the positive ones were slightly larger but not fixed. Note that due to the tedious process of manual labeling, there existed a considerable part of SRCs left unlabeled in the positive set, which may introduce noises to some tasks. Besides, DigestPath contained a still unavailable test set with 227 images from 56 patients, in which 27 images from 11 patients were positive ones. All H\&E stained WSIs involving in the DigestPath dataset were captured at 40$\times$ magnification.
\subsubsection{TCGA dataset}
TCGA program was first launched in 2006 by the National Cancer Institute (NCI) and the National Human Genome Research Institute (NHGRI) of the US. It aimed to comprehensively study the genomic alterations in all major cancers to make outstanding contributions to cancer prevention, diagnosis and treatment. Over the past ten years, more than 20,000 biological samples were collected, covering 33 cancer types with 10 rare ones. TCGA provided abundant representative H\&E diagnostic WSIs. The TCGA Stomach Adenocarcinoma (STAD), Colon Adenocarcinoma (COAD) and Rectum Adenocarcinoma (READ) were popular cohorts in SRC-related classification tasks. On the basis of these data, novel deep learning models to classify the tissue / non-tissue, differentiation grades, and sub-types of tumors were developed \citep{AF2, AF7, AF8, AF9, B2316}. Although we could find SRC related cases in different subsets of TCGA, there was still a lack of systematic integration of SRC cases.

\subsection{Challenges of automatic SRC diagnosis}
\label{subsec:challenges}
\subsubsection{Biological characteristics}
Different from normal cells grown in an organized manner forming specific structures, SRCs are low adhesive with uncertain distribution. Histopathologically, SRCs may be clustered or isolated. At low magnification, the isolated SRCs are easily missed by pathologists and algorithms due to the small size. At high magnification, unlike most other tumor cells with convex shapes, the nuclei of SRCs are squeezed into an irregular crescent shape by intracellular proteins. The uncertainty of the squeezing power leads to the diversity of nucleus morphology, and the flooding of intracellular proteins affects the accurate discrimination of cell boundaries. In addition, SRCs are easily confused with ice crystal bubbles in intraoperative frozen slices. This leads to the difficulty in evaluating whether there are SRCs in the surgical margin tissues or whether there are SRCs metastasis in the omentum and other tissues. Therefore, the recognition of SRCs is challenging for both pathologists and algorithms at multi-scale magnification. 
\subsubsection{Image quality}
Different acquisition equipment and conditions have a great impact on the quality of H\&E stained WSIs, CT, MRI and endoscopic images. Take the H\&E stained WSIs which are usually adopted to confirm the existence of SRCs as examples, the nuclei and cytoplasm are stained into blue and red, and the intracellular proteins are hardly stained. On the one hand, the color distribution of images will be greatly affected by the dyeing process of different batches, concentrations and laboratory environments. On the other hand, the angle and thickness of tissue segmentation, the rupture and overlap of tissue, and other external factors will affect the appearance of cells. In addition, different digital scanners will show different brightness and saturation for the same WSI. Parameters such as focal length during scanning will affect the clarity of WSIs. Therefore, these uncontrolled factors introduce a lot of noises to the SRC image distribution, which puts forward high requirements for the robustness of AI algorithms.
\subsubsection{Manual annotations}
Accurate annotations are of great benefit to the accuracy improvement of AI algorithms, while rough annotations are bound to restrict the performance of models due to noise interference. In fact, since SRCs are difficult even for experienced pathologists to identify without omission, the manual labeling of SRCs requires high professionalism. In addition, pixel-wise labeling is time-consuming and labor-intensive, which further limits the scale of accurate annotations. Usually, pathologists only annotate the cells that are confirmed to be the positive SRC samples, which leads to incomplete labeling in the images. In this situation, if all unmarked areas are regarded as non-SRCs, a large number of SRC noises will be mixed into the negative samples, which will confuse the convergence of the models. In most cases of practical applications, medical centers can only provide patient-level or image-level annotations for algorithm learning, which limits the use of high-performance fully supervised algorithms and weakly supervised algorithms like multiple instance learning (MIL) \citep{MIL} are required.
\subsubsection{Sample imbalance}
The data distribution has a great influence on the model fitting of deep learning. When the quantities of positive and negative samples in the training set are unbalanced, the model will tend to predict the tested samples to be the category with more training samples, so as to minimize the value of the loss function. When the data distribution of the test set is similar to that of the training set, although the biased the model will make the overall evaluation metrics look satisfactory, it will sacrifice the accuracy of few-shot categories, resulting in an intolerable problem of missed detection in clinical practice. In fact, sample imbalance in SRC automatic diagnosis task is common. First, the patient-level samples are imbalanced, since SRC-related patients are far fewer than normal people. Second, the cell distributions are imbalanced. In a screening image, SRCs account for a limited proportion of tissues, while other cells occupy a large area. Third, although SRCs have typical morphological characteristics, the differences among patients, organs, and the image collection process will still introduce disturbances, leading to uncertainty and imbalance of the difficulty of intra class samples. Therefore, to ensure the fairness of the model in SRC automatic diagnosis, the sampling strategy and data augmentation were usually introduced to control the number of samples of different categories in the training set to be similar. In addition, some loss functions were specially designed to increase the loss weights of few-shot samples to force the model to focus on the SRCs. In natural image processing, the long-tailed learning \citep{LongTail} was specially proposed to focus on the widespread sample imbalance issue. However, long-tailed learning has not been embedded in the existed SRC-related studies, since SRC automatic diagnosis task involved fewer categories than natural image processing.

\section{Unimodal algorithms}
\label{sec:unimodal}
\Rone{This section provides a comprehensive overview of unimodal computational algorithms for SRC diagnosis, encompassing image-based, omics-based, and text-based methodologies.} Current research primarily leverages four imaging modalities: H\&E stained histopathology, endoscopic, CT, and MRI data. Following modality-specific pre-processing, AI algorithms extract discriminative features through three principal computational paradigms: classification, detection, and segmentation. \Rone{Notably, vision foundation models offer transferable representations that enhance performance across these downstream tasks.}

\subsection{Image pre-processing}
\label{subsec:pre-processing}
Data pre-processing is commonly required for CNNs and Transformers to extract robust features, which significantly affects the performance of the models. Considering the variance of purposes, we summarize the pre-processing methods into the categories including image normalization, denoising, foreground or regions of interest (RoIs) extraction, data augmentation, and others. The details of pre-processing for each article are listed in Table \ref{table:Pre}.

\onecolumn
\footnotesize
\begin{landscape}
\begin{longtable}{|m{2.0cm}<{\centering}|m{0.75cm}<{\centering}|m{1.1cm}<{\centering}|m{1.5cm}<{\centering}|m{2.25cm}<{\centering}|m{1.75cm}<{\centering}|m{2.5cm}<{\centering}|m{4.0cm}<{\centering}|m{2.5cm}<{\centering}|}
\caption{\label{table:Pre}Summary of data pre-processing and augmentation for SRC-related algorithms (Only the pre-processing mentioned in the articles are included).} \\
	
\hline 
	


\multirow{2}{*}[-7.75ex]{\textbf{Publication}} & \multirow{2}{*}[-7.75ex]{\textbf{Year}} & \multirow{2}{*}[-7.75ex]{\textbf{Task}} & \multirow{2}{*}[-7.75ex]{\textbf{Modality}}  &   \multicolumn{5}{c|}{\textbf{Pre-processing}}  \\ \cline{5-9} ~ & ~ & ~ & ~ &  \textbf{Image normalization} & {\textbf{Denoising}} & \textbf{Foreground (RoI) extraction \& background removal} & \textbf{Data augmentation} & \textbf{Others} \\ \hline

\endfirsthead

\hline 
	
\multirow{2}{*}[-7.75ex]{\textbf{Publication}} & \multirow{2}{*}[-7.75ex]{\textbf{Year}} & \multirow{2}{*}[-7.75ex]{\textbf{Task}} & \multirow{2}{*}[-7.75ex]{\textbf{Modality}}  &   \multicolumn{5}{c|}{\textbf{Pre-processing}}  \\ \cline{5-9} ~ & ~ & ~ & ~ &  \textbf{Image normalization} & {\textbf{Denoising}} & \textbf{Foreground (RoI) extraction \& background removal} & \textbf{Data augmentation} & \textbf{Others} \\ \hline
	
	\endhead
	
	\endfoot
	
	\endlastfoot
	
		
		\citep{AA4} & \makecell[t]{2019} &  Classi-fication & Endo-scopic images & ~ Mean value (from ImageNet dataset) subtraction & \makecell[t]{/} & \makecell[t]{/} & Random flipping, small rotation, elastic deformation, kernel erosions, and dilations  & \makecell[t]{/} \\ \hline
		
		\citep{AL4} & 2019 & Classi-fication & H\&E & / & / & / &  Slight shear / zoom, flipping, whitening & / \\ \hline

		\citep{AF2} & 2020 & Classi-fication & H\&E & / & Gaussian blur smoothing & Thresholding on RGB values & / &  / \\ \hline
		
		\citep{AD4} & 2020 & Object detection & H\&E & / & / & / & Color jitters, horizontal and vertical flipping & / \\ \hline
		
		\citep{AD5} & 2020 & Object detection & H\&E & / & / & / & Random flipping & / \\ \hline
		
		\citep{AF6} & 2020 & Segmen-tation & H\&E & / & / & Otsu thresholding on grayscale image & Random rotations by $90\,{^\circ}$, $180\,{^\circ}$, and $270\,{^\circ}$, flipping, Gaussian and motion blurs, color jitters & / \\ \hline
		
		\citep{AA3} & 2021 & Classi-fication & MRI & Z-score & / & / & / & / \\ \hline
		
		\citep{AF4} & 2021 & Classi-fication & H\&E & / & / & Otsu thresholding on a grayscale version  & / &  / \\ \hline
		
		\citep{AF1} & 2021 & Classi-fication & H\&E & Stain normalization & / & / & CIELAB color space augmentation &  / \\ \hline
		
		\citep{AW1} & 2021 & Classi-fication & H\&E & / & / & / & Flipping, random rotation,  translation, contrast, brightness, hue, saturation & / \\ \hline
		
		\citep{AF7} & 2021 & Classi-fication & H\&E & Color normalization & / & CNN-based classifier & Random rotations by $90\,{^\circ}$, random horizontal and vertical flipping & / \\ \hline
		
		\citep{AF9} & 2021 & Classi-fication & H\&E & Color normalization & / & CNN-based classifier & Random rotations by $90\,{^\circ}$, random flipping, perturbation of the contrast and brightness & / \\ \hline
		
		\citep{AL1} & 2021 & Classi-fication & H\&E & / & / & Otsu thresholding & Filpping, $90\,{^\circ}$ rotations, translations, color shifts & / \\ \hline
		
		\citep{AL2} & 2021 & Classi-fication & H\&E & / & / & / & Rotations of $90\,{^\circ}$, $180\,{^\circ}$, $270\,{^\circ}$ & / \\ \hline
		
		\citep{AL3} & 2022 & Classi-fication & H\&E & Stain normalization & Median blur, Gaussian blur & SLIC & Random rotation &  /  \\ \hline
		
		\citep{AF8} & 2021 & Classi-fication \& segmentation & H\&E & Macenko stain normalization & / & Foreground segmentation based on U-Net & / &  / \\ \hline
		
		\citep{AF10} & 2021 & Classi-fication \& segmentation & H\&E & / & / & Color deconvolution and Otsu thresholding & Flipping, rotation, random cropping and resizing, changes of the aspect ratio and image contrast, Gaussian noise &  / \\ \hline
		
		\citep{AD3} & 2021 & Object detection & H\&E & / & / & / & Flipping and rotation & / \\ \hline
		
		\citep{AD6} & 2021 & Object detection & H\&E & / & / & / & Random flipping & / \\ \hline
		
		\citep{AD7} & 2021 & Object detection & H\&E & / & / & / & Color jitters, flipping & /\\ \hline
		
		\citep{AD1} & 2021 & Object detection & H\&E & / & / & / & Random flipping, data normalization (with a label correction model)& USRNet  (to obtain low-resolution images) \\ \hline
		
		\citep{AA1} & 2022 & Segmen-tation & CT & Z-score & / & / & Random rotation, flipping, cropping &  Resampled with isotropic spacing, clipped intensity values, wavelet filtering\\ \hline
		
		\citep{AS2} & 2022 & Segmen-tation & H\&E & / & / & A coarse U-Net & / & / \\ \hline

        \cite{B2311}&2022&Classi-fication&H\&E&/&/&Otsu thresholding&Brightness, contrast, hue, and saturation modified, JPEG artifacts&/ \\\hline
        \citep{B2316}&2022&Classi-fication&H\&E&/&/&/&Hue and saturation shifts, flipping,
rotation, and random erasing&/ \\\hline
        \citep{B2343}&2022&Classi-fication&H\&E&/&/&/&Flipping, rotation, color changes, and blurring&/ \\\hline
        \citep{B2364}&2022&Classi-fication \& segmentation&H\&E&Color normalization&/&Otsu thresholding in H and S channels of HSV color space&Random cropping, rotation, flipping, and color changes&/\\\hline
        \citep{B2350}&2022&Segmen-tation&H\&E&/&/&/&Changes of 30\% for hue, 0.4 to 1.6 for saturation, 0.7 to 1.3 for brightness, and 0.4 to 1.6 for random scaling / contrast&/ \\\hline
        \citep{B2375}&2023&Classi-fication&H\&E&Stain normalization&/&/&Flipping, rotation, scaling, color jitters, Gaussian blurring and solarization&/ \\\hline
        \citep{B2385}&2023&Classi-fication&H\&E&/&/&DeepLab v3+&Flipping, rotation, scale shifts, brightness, contrast, hue, saturation, Gaussian noises, elastic transforms, grid and optical distortions&/\\\hline
        \cite{J20250402}&2023&Classi-fication&H\&E&Color normalization&/&Automatic detection&Resizing, random horizontal and vertical flipping, and rotation&/\\\hline
\end{longtable}
\end{landscape}
\normalsize

\subsubsection{Image normalization}
Digital images usually have appearance differences such as intensity and color etc., thereby the accuracy of the models will decrease because of insufficient generalization ability. Image normalization is usually employed to eliminate this uncontrolled variability. It emphasizes the real discriminative features without excessive interference from external factors, and accelerates network convergence. For example, Yoon et al. performed the mean value subtraction on each channel of the endoscopic images \citep{AA4}. In addition, z-score normalization is usually utilized with the corresponding mean value and standard deviation to achieve standard normal distributions \citep{AA3, AA1}. Notably, stain / color normalization is relatively necessary for the H\&E images on account of the huge variation in stains, operators and scanner specifications. Its basic principle is to standardize the color appearance of all images to a reference image chosen by an experienced pathologist \citep{stain-Khan, stain-Macenko, stain-Reinhard, stain-Vahadane}. Many SRC-related methods based on H\&E stained images adopted stain / color normalization to enhance the robustness of models to the diversity of staining \citep{AF1, AF7, AF8, AF9, AL3, B2364, B2375,J20250402}.
\subsubsection{Denoising}
Noise is inevitably embedded in the raw images. For instance, air bubbles, compression artifacts, pen marks, blurring, tissue tears and folds, and over-stained areas in H\&E stained samples are irrelevant to the characteristics of SRC lesions, which should be removed to promote the quality of the inputs. Filtering algorithms can reduce the interference of some above factors. Among them, Gaussian blur \citep{AF2, AL3} and median blur \citep{AL3} filters were commonly performed to denoise without severely blurring edges of the objects.
\subsubsection{Foreground (RoI) extraction/ background removal}
In H\&E stained WSIs, object areas containing pathological tissues are usually small, while the blank areas which contribute little to the subsequent tasks occupy the majority. Therefore, the valid parts of WSIs are required to be extracted as the RoIs, and the background needs to be removed. The most direct way to obtain the foreground quickly was to convert the colored images to binary ones \citep{AF2, AF4, AF10, AL1, AF6, B2311, B2364}. It leveraged the difference in grayscale between the foreground and background at low magnification, and identified each pixel through a reasonable threshold. Otsu \citep{Otsu} was the most popular binarization method, which automatically determined the adaptive thresholds by maximizing the inter-class variance in a variety of SRC tasks. In addition, the tissue / non-tissue regions could also be distinguished by CNN-based approaches. For example, CNNs were used as classifiers to choose proper tissue patches for subsequent tumor classification \citep{AF7, AF9}. Besides, segmentation networks such as U-Net \citep{U-Net} and DeepLab v3+ \citep{DeepLabV3Plus} were adopted to outline the foreground \citep{AF8, AS2, B2385}. Compared with the traditional methods based on binarization, the CNN-based methods could flexibly extract specific RoIs according to requirements, but they also introduced additional time cost.
\subsubsection{Data augmentation}
Data augmentation techniques can expand the training set based on existing data, and thus improve the robustness and generalization of models with overfitting minimized. In SRC classification, detection and segmentation, spatial and color transformations were commonly employed. The spatial transformations for SRC diagnosis included horizontal and vertical flipping, rotation, elastic deformation, erosion, dilation, cropping, resizing, and translation \citep{AA4, AF7, AF9, AF10, AL1, AL2, AL3, AL4, AW1, AA1, AF6, AD1, AD3, AD4, AD5, AD6, AD7, B2316, B2343, B2364, B2375, B2385,J20250402}. The color transformations involved CIELAB color space augmentation, superimposed Gaussian noise, whitening, Gaussian blur, motion blur, color shifts, and color jitters including fluctuation of contrast, brightness, hue, and saturation \citep{AF1, AF9, AF10, AL1, AL4, AW1, AF6, AD4, AD7, B2311, B2316, B2343, B2364, B2350, B2375, B2385}. The details of data augmentation in the articles covered in this survey are illustrated in Table \ref{table:Pre}.

\subsubsection{Other pre-processing}
Unlike the common methods mentioned above, some pre-processing operations were related to specific tasks. For example, to obtain the hand-crafted radiomic features, Li et al. resampled each CT image with isotropic spacing in the transverse plane, and then, intensity values were clipped to the range of [-90, 170] to remove outliers \citep{AA1}. Zhang et al. adopted USRNet \citep{USRNet} to reduce the resolution of training data, and the efficiency for SRC detection in low-resolution pathological images was demonstrated \citep{AD1}.

\begin{figure*}[!thb]
	\centering
	\includegraphics[width=\textwidth]{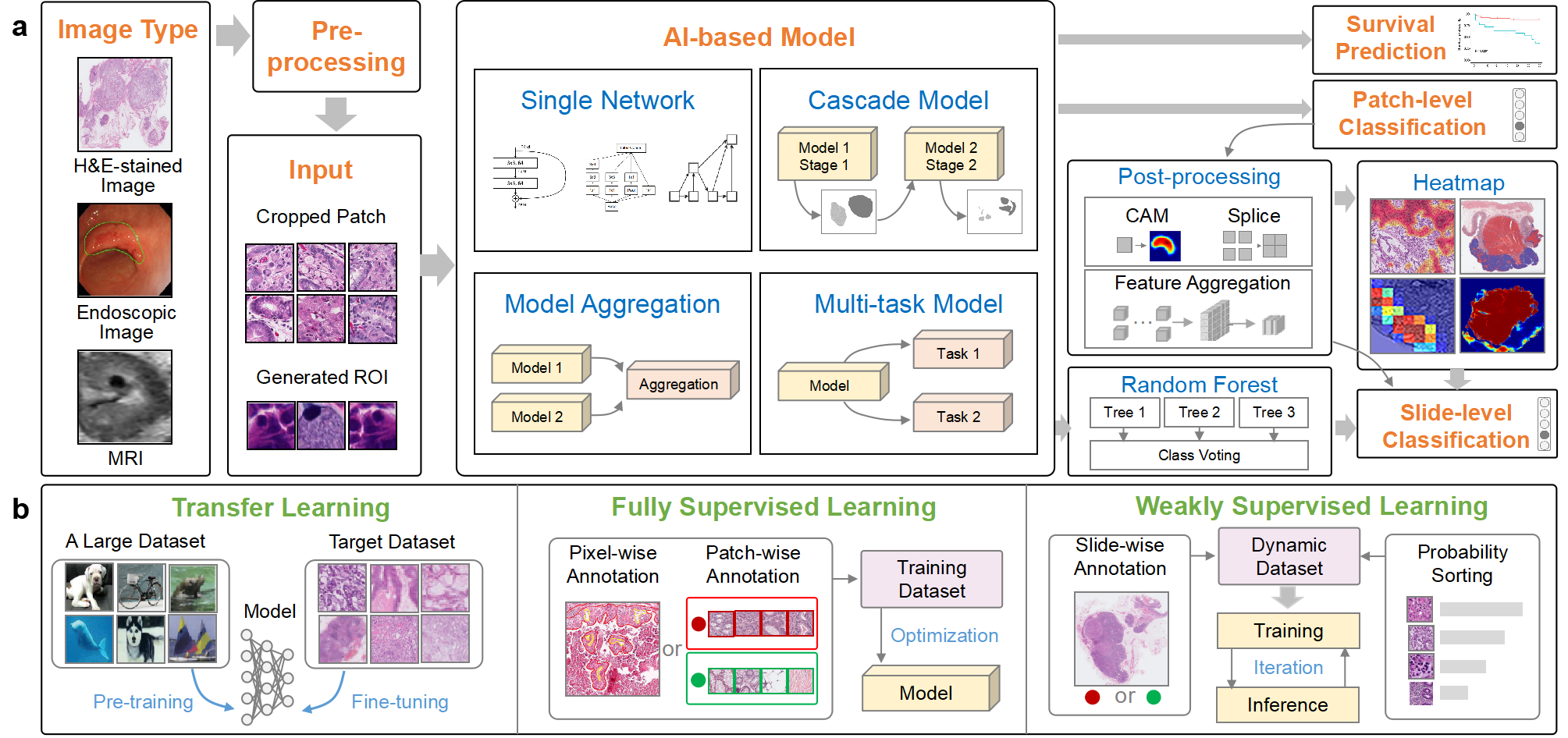}
	\caption{\label{fig:Cla}An overview of classification methods for SRC diagnosis. (a) The process of SRC diagnosis based on classification methods. (b) The learning strategies of SRC classification according to the articles included in this survey.}
\end{figure*}

\subsection{Image classification}
\label{subsec:classification}

Among the classification methods, the target images were fed into the CNNs \Rone{or Transformers} after corresponding pre-processing. As illustrated in Fig. \ref{fig:Cla}, the high-dimensional features of the input images were captured through four types of AI-based classification models. Then the features were adopted to accomplish survival prediction, patch-level prediction, or slide-level prediction. The related models were trained through different combinations of training strategies such as transfer learning, fully supervised learning, and weakly supervised learning. The overview of articles related to SRC classification is summarized in Table \ref{table:Cla}. The details of AI-based models, task-related post-processing, and training strategies are presented next.

\onecolumn

\footnotesize
\begin{landscape}
\begin{longtable}{|m{1.85cm}<{\centering}|m{0.75cm}<{\centering}|m{1.5cm}<{\centering}|m{1.5cm}<{\centering}|m{3.0cm}<{\centering}|m{3.0cm}<{\centering}|m{2.75cm}<{\centering}|m{1.25cm}<{\centering}|m{3.25cm}<{\centering}|}
	\caption{\leftline{\label{table:Cla}Summary of automatic SRC diagnosis algorithms on the basis of classification.}}
 \\
	
	\hline 
	
	\textbf{Publication} & \textbf{Year} & \textbf{Modality} & \textbf{Target} & \textbf{Task} & \textbf{Data} & \textbf{Network} & \textbf{Loss} & \textbf{SRC diagnosis} \\ 
	
	\hline  
	
	\endfirsthead
	
	\hline 
	
		\textbf{Publication} & \textbf{Year} & \textbf{Modality} & \textbf{Target} & \textbf{Task} & \textbf{Data} & \textbf{Network} & \textbf{Loss} & \textbf{SRC diagnosis} \\ 
	
	\hline   
	
	\endhead
	
	\endfoot
	
	\endlastfoot
	
	\citep{AA4} & 2019 & Endo-scopic images & Early gastric cancer & Early gastric cancer detection and depth prediction & 11,539 endoscopic images from 800 patients & VGG-16  & A new loss by adding Grad-CAM  & The histology types of lesions consisted of well / moderately / poorly-differentiated adenocarcinoma, and SRC carcinoma. \\ \hline

	\citep{AA3} & 2021 & MRI & Locally advanced rectal cancer  & Distant metastasis prediction by integrating deep MRI information and clinicopathologic factors & MRI from 235 patients & ResNet-18 & BCE-Loss & SRC carcinoma was part of histologic variants in rectal adenocarcinoma. \\ \hline
	
	\citep{AL4} & 2019 & H\&E & Gastric SRC carcinoma & Histopathologic features of the behavior of gastric SRC carcinoma & 516 images from 10 cases & A 6‐layer CNN & BCE-Loss & SRCs remained within intramucosal areas with poorly differentiated components as dense neighbors. \\ \hline
	
	\citep{AF2} & 2020 & H\&E & Stomach lesion & Identification of well, moderately, and poorly differentiated adenocarcinoma, poorly cohesive carcinoma, and normal gastric mucosa.  & 94 cases of gastroscopic biopsy specimens, and adenocarcinoma WSIs in TCGA (3 stomach and 3 colon cases) & VGG-16, Inception-v3, EfficientNet, MRD-Net \citep{MRD-Net}, N-Net \citep{N-Net}, CAT-Net \citep{CAT-Net}  & MSE-Loss & The method was more accurate in well and moderately differentiated adenocarcinomas than in poorly differentiated adenocarcinomas and poorly cohesive carcinomas including SRCs. \\ \hline
	
	\citep{AL1} & 2021 & H\&E & Gastric SRC carcinoma & Gastric SRC carcinoma classification via fully and weakly supervised learning & 2,824 cases from two hospitals (private, $20\times$ magnification), DigestPath (public, $40\times$ magnification) & EfficientNet-B1& BCE-Loss & The lesions with aggregated SRCs had a high response, while the scattered SRCs had a low response, and the positive probability of a WSI was determined by the highest response of the patches.\\ \hline
	
	\citep{AF1} & 2021 & H\&E & Gastric tumor & Automatical classification into negative for dysplasia, tubular adenoma, or carcinoma based on the method of \citep{Camelyon-liu} & 201 cases of gastric resection and 2,233 cases of biopsy specimens for training, and 7,440 biopsy specimens for evaluation & Inception-v3 for patch classification, an aggregation CNN to generate slide features & CE-Loss & SRC carcinoma was one type of the positive targets. Despite of the overall high accuracy in classifying epithelial tumors, SRC carcinoma suffered from false negatives.  \\ \hline
	
	\citep{AF10} & 2021 & H\&E & Gastric cancer & Screening and localization of gastric cancer based on a multi-task CNN & 10,315 WSIs collected from 4 medical centers & DLA  structure combined with classification and segmentation branches & BCE-Loss & SRC carcinoma was one type of the positive targets.  \\ \hline
	
	\citep{AF3} & 2021 & H\&E & Gastric cancer & Lymph node quantification and metastatic cancer identification & 921 WSIs from 222 patients & Xception, DenseNet-121 & Not mentioned & The system correctly classified most of the patches, but was prone to misdiagnosis when the SRCs were few and scattered. \\ \hline
	
	\citep{AF7} & 2021 & H\&E & Gastric carcinoma & To distinguish differentiated / undifferentiated and non-mucinous / mucinous tumor types & 396 WSIs of 371 patients from TCGA for training, and 232 private WSIs for validation  & Inception-v3 & Not mentioned & SRC carcinoma was regarded as undifferentiated-type. \\ \hline
	
	\citep{AL2} & 2021 & H\&E & Here-ditary diffuse gastric cancer & Detection of regions of hereditary diffuse gastric cancer & 7 gastrectomy specimens (133 annotated tumor foci) & DenseNet-169 & Not mentioned & Regions suspicious for intramucosal SRC carcinoma could be detected. \\ \hline
	
	\citep{AF4} & 2021 & H\&E & Gastric diffuse-type adenocarcinoma & To classify gastric diffuse-type adenocarcinoma from other adenocarcinoma and non-neoplastic subsets & 2,929 endoscopic biopsy cases of human gastric epithelial lesions & EfficientNet-B1, Inception-v3 & CE-Loss & The diffuse-type consisted of poorly-differentiated and SRC carcinoma. \\ \hline
	
	\citep{AF8} & 2021 & H\&E & Colon adenocarcinoma & Tumor microenvironment analysis by recognizing nine different contents & 441 WSIs of 433 patients from TCGA & VGG-19 & Not mentioned & The SRC was not sufficiently presented in the training set and were neglected by the model. \\ \hline
	
	\citep{AF9} & 2021 & H\&E & Color-ectal cancer & Classification of tissue / non-tissue, normal / tumor, and microsatellite stable / instability & 1,920 WSIs from TCGA (colon and rectal cancers) for training, and 365 private WSIs for validation & Inception-v3 & CE-Loss & SRC was associated with the classification of mucinous adenocarcinoma and SRC carcinoma. \\ \hline
	
	\citep{AW1} & 2021 & H\&E & Color-ectal cancer & Identification of nodal micrometastasis based on an annotation-free method \citep{AWR} & 3,182 WSIs from 1,051 patients & ResNet-50 & BCE-Loss & The model performed well on the overall task of identifying micrometastasis and macrometastasis, but slightly worse in identifying SRC and poorly differentiated adenocarcinoma. \\ \hline
	
	\citep{AL3} & 2022 & H\&E & SRC cancer & Detection of ring cell cancer based on RoIs determined by SLIC superpixels & DigestPath dataset & VGG-16, VGG-19, Inception-v3 & Not mentioned & SRC gastric cancer classification was conducted after cropping the small RoIs through SLIC superpixels method.  \\ \hline

    \cite{B2311}&2022&H\&E&Gastric poorly differentiated adenocarcinoma&Poorly differentiated adenocarcinoma classification in gastric endoscopic submucosal dissection with weakly supervised learning&5,103 specimens (2,506 Endoscopic submucosal dissection, 1,866 endoscopic biopsy, and 731 surgical specimen)&EfficientNet-B1&BCE-Loss&SRC carcinoma was included in poorly differentiated adenocarcinoma.\\\hline
    
    \citep{B2343}&2022&H\&E&Gastric cancer&Epstein-Barr Virus (EBV) status prediction from pathology images of gastric cancer biopsy&137,184 patches from 16 tissue microarray (708 tissue cores), 24 WSIs, and 286 biopsy images&ResNet, MobileNet \citep{MobileNet}, EfficientNet, DeiT \citep{DeiT}&CE-Loss&The presence of SRC components was prone to be correlated with gastric cancer specimens without EBV.\\\hline
    
    \citep{B2364}&2022&H\&E&Gastric cancer&Development and multi-institutional validation of an AI-based diagnostic system&984 patients for training and 2,771 patients for validation&GoogLeNet\citep{GoogLeNet}&Not mentioned&SRC could be detected and the performance of a case of poorly cohesive adenocarcinoma with SRCs was discussed.\\\hline
    
    \citep{B2345}&2022&H\&E&Colorectal cancer&Microsatellite instability prediction of colorectal cancer&144 WSIs from 3 hospitals&ResNet, MobileNet, Inception, EfficientNet were embeded in the proposed PPsNet&CE-Loss&Certain histological features like SRC were significantly associated with microsatellite instability.\\\hline
    
    \citep{B2316}&2022&H\&E&Colorectal cancer&DNA mismatch repair (MMR) status prediction based on domain adaption and MIL&441 WSIs from TCGA,78 WSIs from PAIP \citep{PAIP}, and private WSIs (355 from surgical specimens and 341 from biopsy specimens)&DenseNet-121, IBN-net \citep{IBN-net}&Focal Loss, CE-Loss&SRC carcinoma was one of the histology subtypes.\\\hline

    \citep{B2327}&2022&Single-shot femtosecond stimulated Raman scattering&Gastric cancer&Instant diagnosis of gastroscopic biopsy based on single-shot femtosecond stimulated Raman histology&279 patients&Inception-ResNet-V2 \citep{szegedy2017inception}&CE-Loss&Mucinous adenocarcinoma and SRC carcinomas were labeled as undifferentiated cancer in this study.\\\hline
    
    \citep{B2301}&2023&Endoscopic images&Gastric SRC carcinoma&Identification of gastric SRC carcinoma  using few-shot learning&50 gastric benign ulcers, 50 adenocarcinoma and 50 SRC cacinoma&EfficientNetV2-S \citep{EfficientV2}&Not mentioned&Gastric benign ulcers, adenocarcinoma and SRC carcinoma could be classified through K-nearest neighbor classifier based on features from transfer learning.\\\hline
    
    \citep{B2363}&2023&Endoscopic images&Gastric cancer&Cooperation between artificial intelligence and endoscopists for diagnosing invasion depth of early gastric cancer&700 images&EfcientNet-B1&Not mentioned&SRC carcinoma was considered one type of undiferentiated-type cancers.\\\hline
    
    \citep{B2375}&2023&H\&E&Gastric cancer&The development of an AI-based decision support system for gastric cancer treatment&2,440 stomach and 400 colon endoscopic biopsy slides from two hospitals&2-stage multi-scale hybrid ViT \citep{ViT}&CE-Loss&Poorly differentiated tubular, poorly cohesive, SRC, mucinous adenocarcinomas were considered in one class in this study.\\\hline
    
    \citep{B2385}&2023&H\&E&Gastric cancer&Using less annotation workload to establish a pathological auxiliary diagnosis system&1,668 specimens from 1,294 cases&ResNet-50&Not mentioned&AI help pathologists check for easily overlooked SRCs.\\\hline
    
    \citep{B2329}&2023&H\&E&Immune cells and microsatellite instability&The development of a framework for rapid evaluation of CNNs for patch-based histopathology classification&Cropped patches from 6 public datasets&ResNet-18, ResNet-50, ViT&CE-Loss&SRC was one of the histology features of microsatellite instability.\\\hline
    
    \citep{B2328}&2023&H\&E&Colorectal cancer&Lymph node metastasis prediction with weakly supervised learning&843 WSIs from 357 patients&ViT \citep{ViT} embeded in MIL&Not mentioned&Positive lymph nodes in the test set were divided into adenocarcinoma, mucinous carcinoma, and SRC carcinoma subgroups, with the SRC subgroup exhibiting weaker performance.\\\hline

    \citep{J20250402}&2023&H\&E&Colon and gastric cancer& The  discussion of eXplainable AI for CNNs trained to classify microsatellite instability  in colon and gastric cancer&150,078 patches of two classes, 120,063 ones for training and 30,015 ones for validation&Xception\citep{Xception}&Not mentioned&SRC was a subset of known visual features that were indicative of microsatellite instability.\\\hline

    \citep{LongMIL}&2023&H\&E&Stomach, colon and rectal carcinoma&Long-context MIL with attention for survival prediction&TCGA-STAD (321 cases), and TCGA-COADREAD (316 cases)&Transformer optimized by attention with linear bias \citep{LongMIL-Att1} and flash-attetion \citep{FlashAtt}&Survival CE-Loss \citep{SurLoss}&SRCs are typical cases in stomach, colon and rectal carcinoma of TCGA dataset.\\\hline
\end{longtable}
\end{landscape}
\normalsize

\subsubsection{AI-based models}

\textbf{Single Network.} Classification is a classic problem in computer vision. Since AlexNet \citep{AlexNet} achieved remarkable image classification performance over traditional methods on the ImageNet dataset \citep{ImageNet}, more and more CNN architectures have been constructed. Although the network stacked with six convolutional layers by Mori et al. had the ability to extract the features of SRC images \citep{AL4}, the networks with pretrained parameters were often more popular for high accuracy and stability of the models. CNNs designed for classification usually included fully connected layers, which imposed strict constraints on the size of input image patches. The inputs to most CNNs were cropped image patches after pre-processing. There were also methods to obtain RoIs as inputs through manual selection or algorithm generation. For example, Budak et al. generated possible SRC candidates through Simple Linear Iterative Clustering (SLIC) \citep{SLIC} as RoIs \citep{AL3}. The high-dimensional features of these input patches and RoIs were extracted by CNNs to complete the corresponding SRC diagnosis sub-tasks. Empirically, a typical single CNN could usually capture a large number of effective features. Among the methods based on a single network, VGG (VGG-16 and VGG-19) \citep{VGG} was a commonly used efficient network which stacked 13 or 16 convolutional layers and 3 fully connected layers \citep{AA4, AL3, AF2, AF8}. Although VGG had good performance in most classification sub-tasks, its huge amounts of parameters made the fitting relatively difficult. To reduce parameters and increase non-linearity, Inception-v3 \citep{Inception-v3} achieved outstanding performance in the classification of SRCs with factorized convolutions \citep{AL3, AF2, AF1, AF7, AF4, AF9, B2345}. Besides, it was generally believed that the deeper the network, the better the classification effect. However, as the network deepened, the gradient tended to disappear. In practice, the effect was often poor when the networks are too deep. Therefore, ResNet \citep{ResNet} embedded residual learning with shortcuts was proposed to overcome the degradation. Two forms of ResNet, namely, ResNet-18 and ResNet-50 were adopted as single networks for SRC classification \citep{AA3, AW1, B2343, B2345, B2385, B2329}. Then, DenseNet \citep{DenseNet} was proposed to further facilitate cross-layer information to flow, where DenseNet-121 and DenseNet-169 were used for feature extraction of SRC images \citep{AF3, AL2, B2316}. In addition, lightweight networks Xception \citep{Xception}, MobileNet \citep{MobileNet}, and EfficientNet \citep{EfficientNet} were also used for H\&E stained image classification to improve the speed of SRC inference \citep{AL1, AF2, AF3, AF4, B2311, B2343, B2345, B2363,J20250402}. Furthermore, Vision Transformer (ViT) \citep{ViT} has gained prominence in histopathological image analysis tasks due to its capability to capture long-range dependencies and handle diverse object sizes through an attention mechanism \citep{B2375, B2329, B2328}. In summary, the single network approach was the basis for the subsequent multi-model approaches.

\textbf{Model aggregation.} Although an end-to-end single network could extract plenty of effective features, the selection of the network type was still a relatively complex issue. Due to the difference in the original intention of the network design, each single network often had its own merits in practical applications. To complement the advantages of one another, some methods adopted the ensemble learning \citep{EnsembleLearning}. Specifically, different single networks extracted the features of the input patches independently, and then the outputs of different networks were aggregated, so as to achieve the goal of improving accuracy. For example, Hu et al. combined the features extracted by Xception and DenseNet-121 through concatenation, and then the features were further interwoven by fully connected layers to identify the gastric metastatic cancer \citep{AF3}. In summary, it was usually considered that two heads were better than one.

\textbf{Cascade model.} In addition to end-to-end networks, some methods gradually improved the performance by cascading models. Among them, the final classification was split into multiple sub-goals which were implemented in steps. For example, Inception-v3 was implemented twice by Jang et al. to achieve stepwise differentiation of differentiated / undifferentiated and non-mucinous / mucinous tumor types in gastric cancer tissue \citep{AF7}. In addition, Lee et al. applied three sequential classifiers of tissue / non-tissue, normal / tumor and microsatellite stability / high levels of microsatellite instability \citep{AF9}. Similarly, Lou et al. first trained a tumor / non-tumor classifier, and then proposed the PPsNet to classify the microsatellite instability patches \citep{B2345}. The advantage of the cascade model was that the sub-goals could be achieved individually. Specifically, the model trained for the simple sub-goals in the early stage could first divide the samples into multiple sub-spaces. Then, the samples inside the same subspace were more similar than those in different sub-spaces. Therefore, the early-stage classifiers filtered out massive background information that interfered with the prediction, while the later-stage classifiers only focused on the fine-grained separation of similar samples within the subspace. However, the time-consuming of the cascaded model was approximately equal to the sum of the inference time for each sub-goal. Therefore, the more levels the tasks were divided into, the slower the inference was relatively. In summary, the adoption of cascade models in the clinic required a trade-off between speed and accuracy.

\textbf{Multi-task model.} The mining of auxiliary tasks was beneficial to improve the accuracy and convergence speed of the classification models. Among the academic articles covered in this survey, the classification related to SRC diagnosis was a single-label task. In these methods, the loss functions could only constrain the predicted categories, but not delve into whether the focus of the model was correct. Therefore, the auxiliary tasks could improve the attention of the model to the key regions and endow the interpretability of the classification model by embedding the spatial comprehension. For example, Yoon et al. used the weighted sum of gradient-weighted class activation mapping (Grad-CAM) \citep{Grad-CAM} for measuring the localization errors to adjust the classification attention \citep{AA4}. Similarly, Yu et al. embedded both classification and segmentation branches to accomplish gastric cancer screening \citep{AF10}. In addition, Kosaraju et al. proposed a Deep-Hipo structure consisting of a two-stream network with two patches of different scales as input to expand the fields of view \citep{AF2}. In summary, the assistant of the spatial information comprehension embedded in the classification model was of benefit to the accuracy improvement of SRC diagnosis.

\subsubsection{Task-related post-processing}
\textbf{Survival prediction.} The features of input images extracted by AI models could be used for survival prediction \citep{AA3}. The samples could be divided into two groups when setting a lifetime threshold. Then, an AI-based model could be trained to predict the probability of a patient surviving beyond the time threshold only based on his screening images. This probability could be used as an important reference for survival prediction.

\textbf{Patch-level classification.} The AI-based models described above converged through the constraints of the loss functions to extract the salient features of the $i$th input image patch. When the last layer of the network was a fully connected layer, the output was a vector $\vec{P_i}=<p_1^{\left(i\right)},\ p_2^{\left(i\right)},\cdots,p_c^{\left(i\right)}>,$
where $c$ was the number of categories to be distinguished in practical applications. The vector $\vec{P_i}$ was normalized by the softmax function into
\begin{equation}
	\left\{\begin{array}{c}
		\overrightarrow{P_{i}^{\prime}}=<p_{1}^{\prime(i)}, p_{2}^{\prime(i)}, \cdots, p_{c}^{\prime(i)}> \\
		p_{j}^{\prime(i)}=\frac{e^{p_{j}^{(i)}}}{\sum_{k=1}^{c} e^{p_{k}^{(i)}}}, j \in[1, c] \\
		\sum_{j=1}^{c}{p^{\prime}}_{j}^{(i)}=1
	\end{array}\right. , 
\end{equation}
where each element value represented the confidence probability of the corresponding category. Then, the diagnostic patch-level prediction of the input image patch was determined by the category with the max probability. When the last layer only outputs a single value $P_i$ through convolution, the model was usually used to predict the positive probability of the input image patch in a binary classification task. The positive probability was usually normalized by a sigmoid function into 
\begin{equation}
	P_{i}^{\prime}=\frac{1}{1+e^{-P_{i}}}, \quad P_{i}^{\prime} \in[0,1].
\end{equation}
Then, the patch-level prediction was positive when the probability was greater than the preset threshold. For multi-class image classification tasks, cross-entropy loss (CE-Loss) $L_{ce}$ was usually used to constrain the optimization of the models \citep{AA3, AA4, AL4, AL1, AF1, AF4, AF9, AW1, B2343, B2345, B2316, B2327, B2375, B2329}, and the loss of each sample could be calculated as
\begin{equation}
	L_{c e}(i)=-\sum_{j=1}^{c} y_{j}^{(i)} \log \left(p_{j}^{\prime(i)}\right),
	\label{equ:ce}
\end{equation}
where $c$ was the number of categories, ${p^\prime}_j^{\left(i\right)}$ was the normalized predicted probability, and $y_j^{\left(i\right)}$ was the ground truth. $y_j^{\left(i\right)}$ was 1 when $j$ was the annotated true category, and 0 otherwise. When there were only two categories, the input image was either positive or negative. The output vector after normalization could be represented by a vector $<1-P_i^\prime,P_i^\prime>$, where $P_i^\prime$ was the positive probability. Then, CE-Loss was equivalent to binary cross-entropy loss (BCE-Loss):
\begin{equation}
	L_{b c e}(i)=-y^{(i)} \log \left(P_{i}^{\prime}\right)-\left(1-y^{(i)}\right) \log \left(1-P_{i}^{\prime}\right),
	\label{equ:bce}
\end{equation}
which was also popular in the SRC-related diagnosis methods \citep{AA3,AL4,AL1,AF10,AW1,B2311}. The models that output only one value representing the probability of being positive could also be optimized with mean squared error loss (MSE-Loss) \citep{AF2}:
\begin{equation}
	L_{m s e}(i)=\left(P_{i}^{\prime}-y^{(i)}\right)^{2},
\end{equation}
where $P_i^\prime$ was the output probability and $y^{(i)}$ was the ground truth of the input image $i$.

\textbf{Slide-level classification.} A single H\&E stained WSI could crop tens of thousands of patches for AI model input, and thus, a patch-level prediction could only measure the tip of the iceberg. Therefore, the slide-level classification required synthesizing the predictions of all the patches cropped from the target slide. The most direct way was to splice the patch-level predictions according to the cropping index positions, so that the approximate positions of the lesions in the original slide could be clearly point out \citep{AF7, AL1, AF2, AF8, AF9}. However, an image patch only corresponded to a single output category, which did not clarify the location of key regions in the input patch. In addition, one patch-level prediction only represented one pixel in the spliced heatmap corresponding to the H\&E stained WSI, resulting in the spliced diagnostic heatmap being $s$ times of down-sampling of the original slide, where $s$ was the stride for cropping the patches. Yoon et al. adopted Grad-CAM to decide the importance of each neuron in the last convolutional layer by gradients, thereby prompting the classification diagnosis of the endoscopic images to originate from the correct mining of the lesion locations \citep{AA4}. Similarly, Kanavati et al. obtained larger and smoother heatmaps by Grad-CAM than direct splicing \citep{AL1}. Clinically, not only diagnostic heatmaps were expected, but also slide-level diagnosis which required comprehensive measurement of heatmaps in a quantitative manner. Park et al. trained a random forest classifier with the extracted features to obtain the slide-level categories of gastric WSIs \citep{AF1}. Random forest classifier could not only be trained with features extracted by AI models, but also could be embedded with many artificially designed features such as number of connected domains, maximum positive probability, mean positive probability of the target slide. The importance of these features in this task could also be measured meanwhile. However, artificially designed features were sometimes too subjective and required much practical experience. Drawing on the ideas of MIL, after training the patch-level classification model, they removed the fully connected layers and spliced the high-dimensional features output by the last convolutional layer of all corresponding patches according to the cropping positions, and then trained a cascade network to get the slide-level classification automatically.

\subsubsection{Training strategies}
\textbf{Transfer learning.} Deep learning is a data-driven machine learning method. Generally, large-scale high-quality training data are essential for AI models. However, due to multiple factors such as privacy and difficulty in labeling, large-scale medical images for training are difficult to achieve. Therefore, transfer learning is an efficient and effective approach for AI-based medical image processing. Among the SRC-related articles, when using classic network structures as the backbones, the parameters pretrained on the large-scale ImageNet dataset \citep{ImageNet} were usually used as the initialization parameters \citep{AA3,AA4,AL1,AL3,AF1, AF2, AF3, AF4, AF7, AF8, AF9, AF10, AW1, B2345, B2363}. The initialized models were then fine-tuned on the target task. Transfer learning accelerated model convergence while avoiding the model falling into local optimal fitting. In summary, transfer learning often played an important role in the initial stage of training.

\textbf{Fully supervised learning.} Fully supervised learning was the most common classification training strategy in the articles covered in this survey. Among them, each input image patch was equipped with an accurate and unique ground-truth label. For example, in the tasks dedicated to SRC recognition, patches included SRCs were usually labeled 1, and 0 otherwise \citep{AA4, AL4, AL1, AL3}. Similarly, other algorithms for fully supervised classification assigned different labels to the target categories \citep{AA3, AF1, AF10, AF3, AF7, AL2, AF4, AF8, AF9, B2343, B2364, B2345, B2329, B2363}. The output of the AI models were the confidence probabilities that the input patches belonged to a certain category. Since each input patch had a clear true label, ideally only the predicted probability corresponding to the truth should be 1, while that of the wrong categories should be 0. The model iteratively optimized the parameters by back propagation to minimize the difference between the predicted and ideal probabilities. Therefore, when the model converged, the classification model of fully supervised learning had moved close to the ideal state as much as possible, namely, the category with the highest confidence probability was more likely to be the true diagnostic category for the input image patch.

\textbf{Weakly supervised learning.} Fully supervised learning often requires careful annotations for giga-resolution slides which is labor-intensive and time-consuming. On the contrary, data with slide-level annotations are less difficult to obtain. To take advantage of the large-scale data with weak labels, weakly supervised learning is a good choice. However, for giga-resolution H\&E stained slides, with the slide-level labels, the location of lesions cannot be determined directly. The biggest challenge of weakly supervised learning is that lesions may only account for a very small part, so a large number of mislabels will be introduced if slide-level labels are directly assigned to each cropped image patch. Therefore, MIL, a typical weakly supervised learning method, was applied to SRC diagnosis \citep{AL1, AF4, AW1,B2311,B2328,B2316,B2375,LongMIL}. MIL first loaded the patches cropped from a slide into a bag, and then assigned the slide-level label to the bag. If the bag label was negative, then all patches in it were negative. If the bag label was positive, at least one patch in the bag was positive, but the label of each patch was not specified. For example, a small amount of fully supervised data were first used to train the initialization model which was further optimized though selecting suitable training data through iterations \citep{AL1, AF4,B2311,B2316}. Specifically, the fully supervised training model inferred the data in the bags, and added the top $k$ patches with the highest confidence probabilities consistent with the bag label to the training set to retrain the model, where $k$ was a hyper-parameter. In this way, training and inference were continuously alternated, and the accuracy of the model was gradually improved. Additionally, due to the alignment between the continuous cropping of WSI patches and ViT's input pattern, coupled with ViT's attention mechanism that effectively captures crucial positive regions, encoding patch features through ViT allows for the acquisition of slide-level classification results \citep{B2328,B2375}. In summary, weakly supervised learning was an effective way to mine high representational information in SRC data. 

\subsection{Image detection}
\label{subsec:detection}

\begin{figure*}[!th]
	\centering
	\includegraphics[width=\textwidth]{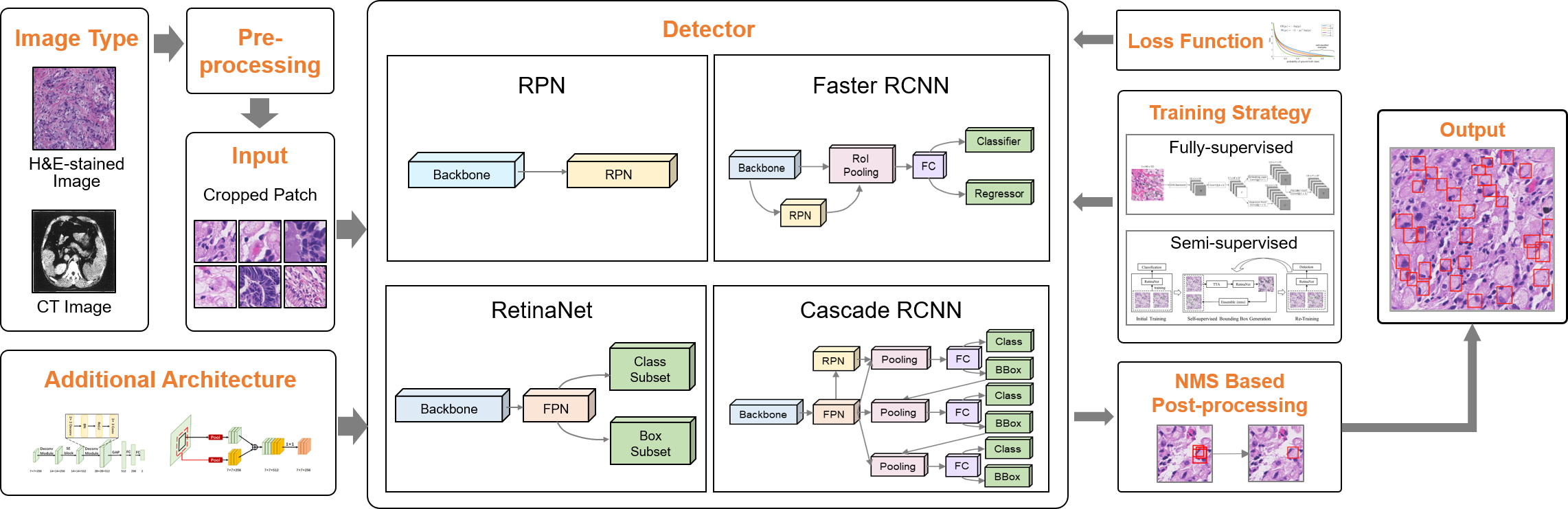}
	\caption{\label{fig:Det}An overview of detection methods for SRC diagnosis.}
\end{figure*}

The task of object detection is to find all objects of interest in an image and synchronously determine their categories and locations. Fig. \ref{fig:Det} illustrates the main pipeline of SRC detection. Specifically, four typical detectors based on various backbone networks were usually adopted as well as some other architectures. Different novel loss functions were elaborately designed to constrain the fitting and convergence of the models. Although the training strategies were sometimes different, the final outputs were usually generated by Non-Maximum Suppression (NMS) based post-processing to remove overlapping bounding boxes. Details of SRC detection methods are illustrated in Table \ref{table:Det}. To clarify, we divide the whole procedure into four parts: backbone, detector, loss function and training strategy.

\onecolumn
\footnotesize
\begin{landscape} 
\begin{longtable}{|m{1.85cm}<{\centering}|m{0.7cm}<{\centering}|m{1.5cm}<{\centering}|m{1.5cm}<{\centering}|m{2.75cm}<{\centering}|m{1.25cm}<{\centering}|m{2.5cm}<{\centering}|m{2.5cm}<{\centering}|m{3.0cm}<{\centering}|}
	\caption{\leftline{\label{table:Det}Summary of automatic SRC diagnosis algorithms on the basis of object detection.}}
 \\
	
	\hline 
	
	\textbf{Publication} & \textbf{Year} & \textbf{Modality} & \textbf{Target} & \textbf{Data} & \textbf{Method type} & \textbf{Network} & \textbf{Loss} & \textbf{SRC diagnosis} \\ 
	
	\hline  
	
	\endfirsthead
	
	\hline 
	
		\textbf{Publication} & \textbf{Year} & \textbf{Modality} & \textbf{Target} & \textbf{Data} & \textbf{Method type} & \textbf{Network} & \textbf{Loss} & \textbf{SRC diagnosis} \\ 
	
	\hline   
	
	\endhead
	
	\endfoot
	
	\endlastfoot
		\hline
		\citep{AA2} & 2019 & CT & Peri-gastric meta-static lymph nodes & Initial group: 18,780 enhanced CT images and 1,371 labeled CT images from 313 patients 
		\newline Precision group: 11,340 enhanced CT images and 1,004 labeled CT images from 189 patients \newline
		Verification group: 6,000 CT images from 100 patients & Two-stage, anchor based & FR-CNN(Faster-RCNN, backnone: VGG-16) & / & The differentiation level of gastric cancer consisted of well / intermediate / poor / SRC carcinoma. \\ \hline
		
		\citep{AD4} & 2020 & H\&E & SRCs & Digestpath dataset & One / two-stage, anchor based & RPN, Faster RCNN, RetinaNet & CE-Loss, Smooth L1 Loss, Triplet Loss & A similarity learning approach for SRC detection. \\ \hline
		
		\citep{AD5} & 2020 & H\&E & SRCs & Digestpath dataset & One / two-stage, anchor based & Cascade RCNN (backbone: ResNet-50 with FPN) & Smooth L1 Loss and CE-Loss & SRC detection with classification reinforcement 
		detection network. \\ \hline
		
		\citep{AD6} & 2021 & H\&E & SRCs & Digestpath dataset & One / two-stage, anchor based & Cascade RCNN (backbone: ResNet-50 with FPN) & Smooth L1 Loss and CE-Loss & SRC detection with classification reinforcement 
		detection network. \\ \hline
		
		\citep{AD1} & 2021 & H\&E & SRCs & DigestPath dataset for training and validation, a private dataset for test & One-stage, anchor based & USRNet, RetinaNet, Label correction model(classi-fier, backbone: ResNet-18) & RGHMC Loss & The framework provided an essential method for SRC detection in low-resolution pathological images. \\ \hline
		
		\citep{AD2} & 2021 & H\&E & SRCs & DigestPath and MoNuSeg \citep{MoNuSeg} dataset & One-stage, anchor based & Two RetinaNets (backbone: ResNet-18) for classification and detection respectively & Focal Loss for classification, Smooth L1 Loss for box regression & A semi-supervised deep convolutional framework for SRC detection to deal with the issue of incomplete annotations. \\ \hline
		
		\citep{AD3} & 2021 & H\&E & SRCs & Digestpath dataset & One-stage, anchor based & RetinaNet  (backbone: ResNet-18) & DGHM-C Loss & A novel DGHM-C Loss was proposed for partially annotated SRCs detection. \\ \hline
		
		\citep{AD7} & 2021 & H\&E &  Nuclei and SRCs & DigestPath and MoNuSeg dataset & One / Two-stage, anchor based & RPN, Faster RCNN, RetinaNet \newline
		(backbone: ResNet-50 / ResNet-101 / ResNeXt-101) & CE-Loss and Focal Loss for classification, Smooth L1 Loss for regression, Pair Loss and Triplet Loss for embedding & A general similarity-based method for both nuclei and SRCs detection. \\ \hline

        \citep{B2302}\citep{B2302again}&2022&H\&E&SRCs&200 images (Part from two hospitals and the others from internet sources)&Two-stage, anchor based&Fast-RCNN (backbone: ResNet-50)&Smooth L1 Loss, CE-Loss&The SRCs were annotated by bounding boxes and detected by a general detection method.\\\hline
        
        \citep{B2314}&2023&H\&E&SRCs&770 patches with size from 108 WSIs of 9 patients&Two-stage, anchor based&C3Det \citep{C3Det}, Faster-RCNN (backbone: ResNet50)&Not mentioned&An interactive detection method with bounding boxes generated by NuClick \citep{NuClick}.\\\hline
        
        \citep{B2321}&2023&H\&E&SRCs&DigestPath dataset&One-stage, anchor based&RetinaNet&CE-Loss&The learned representation  from multiple H\&E data sources could be used to improve the performance of additional tasks via transfer learning such as SRC detection.\\\hline
        
\end{longtable}
\end{landscape}
\normalsize

\subsubsection{Backbone}
Backbone network was one of the most important components of state-of-the-art (SOTA) detectors. Among them, VGG \citep{VGG}, ResNet \citep{ResNet} and ResNeXt \citep{ResNeXt} were three typical backbone architectures utilized for SRC detection. 

\textbf{VGG.} VGG was proposed on the basis of AlexNet. Differently, VGG used small convolution kernels to increase the network depth. VGG achieved advanced results on ImageNet and became one of the most commonly used backbone networks for image classification and object detection. For SRC detection, VGG16 was a popular choice as the backbone network.

\textbf{ResNet.} ResNet with residual blocks was proposed to alleviate degradation risks. ResNet won the first place in all five main tracks of ILSVRC 2015\footnote{\url{https://image-net.org/challenges/LSVRC/2015/}}  (Accessed July 21, 2025) and MS COCO 2015\footnote{\url{https://cocodataset.org/\#home}}  (Accessed July 21, 2025) competitions, and achieved robust and good performance in many specific tasks. For SRC detection, the ResNet-18, ResNet-50 and ResNet-101 architectures were commonly adopted as the backbone networks.

\textbf{ResNeXt.} ResNeXt was proposed based on ResNet and Inception module \citep{GoogLeNet}, which adopted group convolution modules in residual blocks, namely, the ``split-transform-merge" mode of Inception. Additionally, a simple and efficient architecture was achieved by applying identical topological paths in ResNeXt blocks instead of the elaborate Inception transformation. Cardinality was the only hyperparameter to control the number of convolutional paths, which could be considered as the third dimension of the data to improve the performance in addition to width and depth. Therefore, ResNeXt could achieve higher accuracy while consuming slightly fewer parameters than a similar depth ResNet architecture. ResNeXt won the runner-up of the ILSVRC 2016 challenge \footnote{\url{https://image-net.org/challenges/LSVRC/2016/}}  (Accessed July 21, 2025). The typical ResNeXt-50 and ResNeXt-101 have been applied to the SRC detection.

\subsubsection{Detector}
\begin{figure*}[!t]
	\centering
	\includegraphics[width=\textwidth]{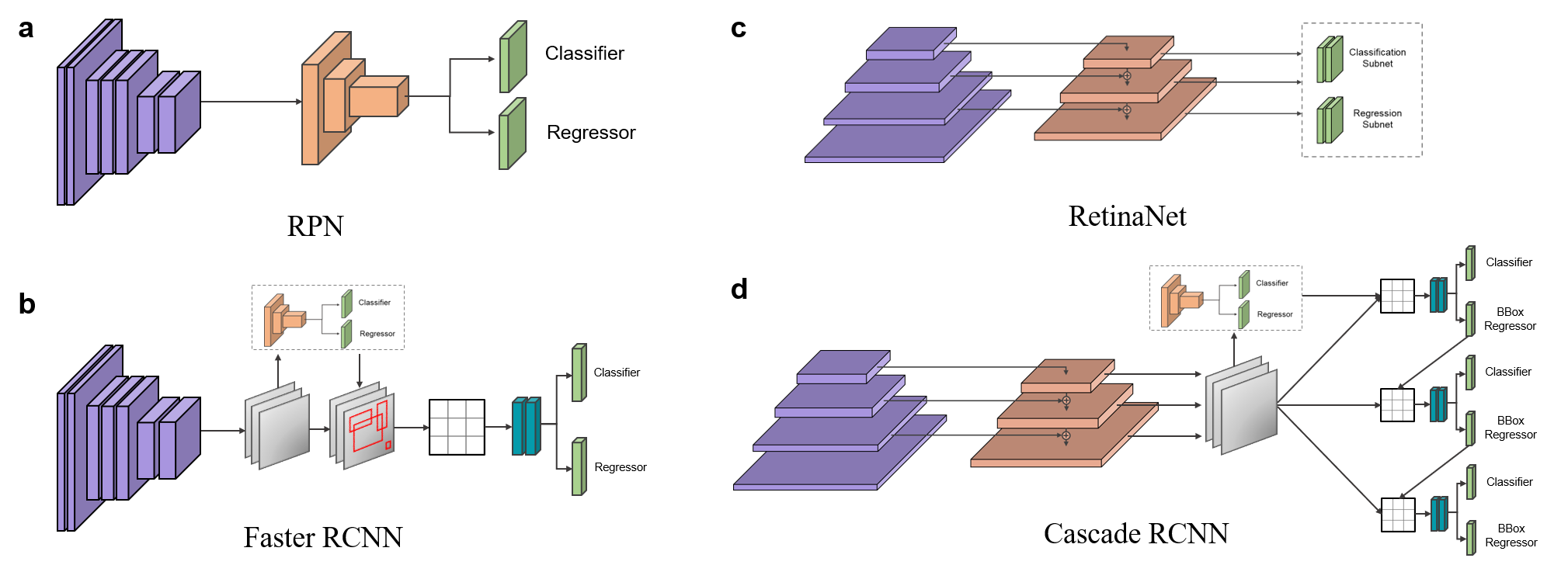}
	\caption{\label{fig:Net}The typical architectures of detectors for SRC detection. (a) RPN \citep{FasterRCNN}. (b) Faster RCNN \citep{FasterRCNN}. (c) RetinaNet \citep{RetinaNet}. (d) Cascade RCNN \citep{CascadeRCNN}.}
\end{figure*}

Detector, namely, the entire object detection network, output the classification and localization results of objects. According to whether and how many region proposal modules were introduced, there were three types of detectors: one-stage (e.g., RetinaNet \citep{RetinaNet}), two-stage (e.g., Faster RCNN \citep{FasterRCNN}) and multi-stage (e.g., Cascade RCNN \citep{CascadeRCNN}) detectors. Among them, four frequently used detectors for SRC detection shown in Fig. \ref{fig:Net} are presented in this section.

\textbf{Region proposal network (RPN).} Region proposal methods and region-based CNNs (RCNNs) were two essential components of the two-stage detectors. Nevertheless, traditional region proposal methods such as Selective Search \citep{SelectiveSearch} and EdgeBoxes \citep{EdgeBoxes} were time-consuming, making it impossible for the detectors to be real-time. The introduction of RPN \citep{FasterRCNN} (Fig. \ref{fig:Net}a) made the process almost cost-free. It took the feature maps from the backbones of an arbitrary image as input and output bounding boxes and corresponding scores of objects. Particularly, anchor boxes were introduced as localization references with various sizes and aspect ratios. Specifically, RPN slided over the feature maps extracted from backbones with a 3$\times$3 convolutional kernel and obtained a 256-dimension feature vector for each location. These feature maps were then fed into two branches: one for classification and another one for regression. The branch for classification generated the confidence probability for each predicted box containing corresponding object, and the branch for regression refined the position and size of each bounding box based on the corresponding anchor. RPN was widely used in current two-stage detection networks, such as Faster RCNN and Cascade RCNN. For SRC detection, it was also served as an independent detection network.

\textbf{Faster RCNN.}  Faster RCNN (Fig. \ref{fig:Net}b) was one of the most popular two-stage detectors. It was essentially a Fast RCNN \citep{Fast-R-CNN} with RPN which was the core novelty as a nearly cost-free proposal algorithm. The backbone network served images as input and generated feature maps which were then fed into RPN to obtain region proposals. Next, the chosen proposals were mapped back to the previous feature maps in RoI Pooling layer and fed into fully connected layers to obtain ultimate classification and regression results. The training process of Faster RCNN contained four steps. First, the backbone network was initialized with the parameters pretrained on the ImageNet dataset and the RPN was fine-tuned end-to-end. Second, Fast RCNN was trained with another pretrained model and rectangular proposals generated by the RPN of last step. Then, only the parameters of RPN were updated while the rest parameters of the model in the second step were fixed. Finally, the parameters of the RPN and the backbone were fixed, and the unique layers of Fast RCNN were fine-tuned. Faster RCNN was the first unified and near real-time object detection framework based on deep learning, and its core principles have inspired many subsequent detectors. In addition, Faster RCNN has also been widely used in SRC detection.

\textbf{RetinaNet.} One-stage detectors were popular for their high speed and simpleness, but their precision was far behind that of two-stage detectors. In two-stage detectors, sparse sampling and NMS algorithm helped filter out most negative samples in the RPN to attain better performance. To alleviate extreme imbalance between foreground and background in dense detection with one-stage detector, Lin et al. proposed a novel classification loss called Focal Loss \citep{RetinaNet} reduce the loss weights of the simple samples and focus on the hard ones. To verify the effectiveness of Focal Loss, Lin et al. also proposed a simple one-stage detector, RetinaNet (Fig. \ref{fig:Net}c), which utilized the Feature Pyramid Network (FPN) \citep{FPN} to extract and fuse features of different layers from the backbone, and used two similar sub-nets at each layer for classification and regression, respectively. One the one hand, detection of multi-scale objects was implemented at different levels of feature maps. One the other hand, features with high-level semantic information and high resolution were integrated, which was beneficial for classification and localization. RetinaNet surpassed all the other detectors in both accuracy and speed when it was proposed, so it has been widely used in SRC detection.

\textbf{ Cascade RCNN.} In the training process of RPN, an Intersection and union (IoU) threshold was defined to classify positive and negative examples. A detector tended to generate noises with a relative low threshold, while suffered a performance degradation with increasing the IoU threshold. Two main reasons for this problem were considered: one was the overfitting due to rapid reduction of positives, and another one was the mismatch of the proposal quality between training and test. Cai et al. demonstrated that regressors trained with different IoU thresholds could provide the best optimization for samples of IoU close to the corresponding thresholds \citep{CascadeRCNN}. To this end, a multi-stage object detection framework, Cascade RCNN (Fig. \ref{fig:Net}d) was proposed to gradually improve the quality of the bounding boxes to alleviate the problems of overfitting in the training process and quality mismatch in the inference process. Specifically, three detectors were trained in a cascaded way with IoU thresholds of {0.5, 0.6, 0.7}, while each detector adopted the optimized bounding boxes from the previous stage, thereby refining the proposals step by step. As a result, sufficient positives were produced for each stage to prevent overfitting, and cascaded optimization could also vanish the mismatch. Cascade RCNN was shown to be applicable to a wide range of object detection architectures, and it also appeared in the SRC detection task.

\subsubsection{Loss functions for detection}
The SRC detection task aims at generating an accurate bounding box for each SRC in the input images. According to different requirements, different novel loss functions were proposed to constrain the convergence of the models. Among them, classification loss functions were used to constrain the models to generate bounding boxes around SRCs, while regression loss functions were used to calibrate the position of the bounding boxes. Classification loss, regression loss and loss functions for other special purposes used in the SRC detection methods are described in this section.

Firstly, the classification loss functions for SRC detection are illustrated as follows. 

\textbf{CE-Loss.} CE-Loss was exactly the most popular choice for classification tasks, which was also adopted in the classification branches of the detection models. The definition of CE-Loss used in SRC detection tasks was the same as that in SRC classification tasks (Equation \ref{equ:ce} and Equation \ref{equ:bce}). We redefined the loss function for the convenience of the following description as
\begin{equation}
	L_{c e-d}\left(p^{(i)}, c^{(i)}\right)=\left\{\begin{array}{cl}
		-\log \left(p^{(i)}\right), &c^{(i)}=1 \\
		-\log \left(1-p^{(i)}\right), &c^{(i)}=0
	\end{array}\right. ,
\end{equation}
where $c^{(i)}$ is the one-hot encoded label of the sample $i$, and $p^{(i)}$ represents its predicted probability of the category with $c^{(i)}=1$. The mean value of $L_{ce-d}$ of all examples in a batch was used for back propagation, thereby promoting the model to improve the classification accuracy.

\textbf{Focal Loss} \citep{RetinaNet}. In object detection, dense sampling led to an extreme class imbalance, that is, a vast number of easy negatives would dominate the whole loss and thus overwhelm the training. Focal Loss was proposed to reduce the contribution of easy samples, while focus on hard and misclassified ones. For notation convenience, the gradient norm $g_i$ was introduced to measure the difference between prediction $p^{(i)}$ and the ground-truth $c^{(i)}$ for the sample $i$: 
\begin{equation}
	g_{i}=\left\|p^{(i)}-c^{(i)}\right\|=\left\{\begin{array}{cl}
		1-p^{(i)},& c^{(i)}=1 \\
		p^{(i)}, &c^{(i)}=0
	\end{array}\right. .
\end{equation}
Correspondingly, Focal Loss was formulated as:
\begin{equation}
	L_{focal}=\frac{1}{N} \sum_{i=1}^{N} \alpha g_{i}^{\gamma} L_{ce-d}\left(p^{(i)}, c^{(i)}\right),
\end{equation}
where $\alpha$ balanced the importance between the foreground and background, $g_i^\gamma$ denoted a modulation factor of the $L_{ce-d}$ with a focusing parameter $\gamma\geq0$. Obviously, the relative loss of well-classified examples was down weighted when $\gamma>0$, and Focal Loss degenerated into the standard CE-Loss when $\gamma=0$. Experiments by Lin et al. showed $\gamma=2$ and $\alpha=0.25$ worked best \citep{RetinaNet}.

\textbf{GHMC Loss} \citep{loss-GHMC}. Extremely hard examples were considered as outliers, whose gradient directions tended to vary from others. Thus, models usually got confused when balancing their gradients with Focal Loss. To this end, GHMC Loss was proposed to reduce the contribution of outliers besides well-classified samples. Experiments has demonstrated that samples with gradient norm close to 0 (easy samples) or 1 (difficult samples) occupied a significantly larger proportion than others. To measure the difficulty of a sample $i$, the gradient density $GD$ and the harmonizing parameter $\beta_i$ were defined:
\begin{equation}
	GD\left(g_{i}\right)=\frac{1}{l_{\varepsilon}\left(g_{i}\right)} \sum_{k=1}^{N} \delta_{\varepsilon}\left(g_{k}, g_{i}\right),
	\label{eq:GD}
\end{equation}
\begin{equation}
	\beta_{i}=\frac{N}{G D\left(g_{i}\right)},
\end{equation}
where $g_i$ and $g_k$ represented the gradient norm of the $i$th and $k$th sample, respectively, and $\delta_\varepsilon\left(g_k,g_i\right)$ represented whether $g_k$ was in the range centered on $g_i$ with the valid length $l_\varepsilon\left(g_i\right)$. $l_\varepsilon\left(g_i\right)$ and $\delta_\varepsilon\left(g_k,\ g_i\right)$ were defined as
\begin{equation}
	\text{For }  \forall \varepsilon>0,\left\{\begin{array}{l}
		\begin{aligned}
  l_{\varepsilon}\left(g_{i}\right)=\min &\left(g_{i}+\frac{\varepsilon}{2}, 1\right) \\& -\max \left(g_{i}-\frac{\varepsilon}{2}, 0\right) \end{aligned}\\
		\delta_{\varepsilon}\left(g_{k}, g_{i}\right)=\left\{\begin{array}{ll}
			1, & \text{if } \lvert g_{k}-g_{i}\rvert \leq \frac{\varepsilon}{2} \\
			0, & \text{otherwise} 
		\end{array}\right. .
	\end{array}\right.
\end{equation}
Thus, $GD\left(g_i\right)$ denoted the gradient density around $g_i$, and the parameter $\beta_i$ varied inversely with it, which down-weighted the easy samples and outliers. Then, the GHMC Loss was formulated as
\begin{equation}
	L_{g h m c}=\frac{1}{N} \sum_{i=1}^{N} \beta_{i} L_{c e-d}\left(p^{(i)}, c^{(i)}\right) .
	\label{eq:ghmc}
\end{equation}

\textbf{RGHMC Loss} \citep{AD1}. RGHMC Loss was a modified version of GHMC Loss, which aimed to handle the incomplete annotation problem in the DigestPath dataset. Specifically, a considerable amount of unlabeled SRCs introduced noises to the negative set during training. Therefore, the revised ground-truth $c_r^{(i)}$ was introduced:
\begin{equation}
	c_{r}^{(i)}=\left\{\begin{array}{l}
		1, x \in A_{P} \cup A_{R}, A_{R} \subset A_{N}^{\text {noisy }} \\
		0, x \in A_{N} \backslash A_{R}
	\end{array},\right.
\end{equation}
where $A_P$ and $A_N$ represented sets of original positive and negative samples, respectively, and $A_R$ denoted a recall set from negatives considered as SRCs. The combination of the detection model and an auxiliary classifier was implemented by Zhang et al. \citep{AD1} to determine each element of $A_R$:
\begin{equation}
	A_{R}=\left\{x \in A_{N} \mid l(x)=1, p^{c}(x)>t_{1}, p(x)>t_{2}\right\},
\end{equation}
where $p^c\left(x\right)$ and $l\left(x\right)$ indicated the probability and label predicted by a well-trained auxiliary classifier, respectively, $p\left(x\right)$ was the classification score of the detection network, $t_1$ and $t_2$ were two tunable thresholds. The RGHMC Loss was finally defined as
\begin{equation}
	L_{r g h m c}=\frac{1}{N} \sum_{i=1}^{N} \beta_{i} L_{c e-d}\left(p^{(i)}, c_{r}^{(i)}\right).
	\label{eq:rghmc}
\end{equation}

\textbf{DGHM-C Loss} \citep{AD3}. DGHM-C Loss modified the original GHMC in another way to adapt to partially annotated object detection. Lin et al. \citep{AD3} argued that outliers in clean data space were probably hard samples worth learning, while those in noisy data space were rather likely to be mislabeled. To this end, a novel DGHM strategy was proposed to decouple the noisy samples from the clean ones, and the $GD$ in Equation \ref{eq:GD} was modified separately as
\begin{equation}
	G D\left(g_{i}\right)=\left\{\begin{array}{l}
		\frac{1}{l_{\varepsilon}\left(g_{i}\right)}\left(\sum_{k=1}^{N_{c}} \delta_{\varepsilon}\left(g_{k}, g_{i}\right)\right), x_{i} \in S_{c} \\
		\frac{1}{l_{\varepsilon}\left(g_{i}\right)}\left(\sum_{k=1}^{N_{n}} \delta_{\varepsilon}\left(g_{k}, g_{i}\right)\right), x_{i} \in S_{n}
	\end{array},\right.
\end{equation}
where $S_c$ and $S_n$ represented the clean data space (including annotated positives and all samples of negative images) and the noisy data space (negatives in positive images), respectively, and $N_c$ and $N_n$ were the number of their anchors. Harmonizing parameter $\beta_i$ was also redefined with a modulating factor $\gamma_i$:
\begin{equation}
	\beta_{i}=\frac{N}{G D\left(g_{i}\right)^{\gamma_{i}}},
\end{equation}
\begin{equation}
	\gamma_{i}=\left\{\begin{array}{ll}
		\mu_{n},& g_{i} \geq \lambda, x_{i} \in S_{n} \\
		\mu_{c},& g_{i} \geq \lambda, x_{i} \in S_{c} \\
		1, &\text { otherwise }
	\end{array},\right.
\end{equation}
where $\gamma_i$ had two different values for outliers exceed the threshold $\lambda$ in $S_c$ and $S_n$. In general, $\mu_n\geq1$ was selected to reduce the weight of outliers in $S_n$ to prevent overfitting to noises, and $\mu_c\le1$ was chosen at the same time to up-weight the outliers in $S_c$ to achieve hard sample mining. In SRC detection, $\mu_n=2.0$, $\mu_c=0.5$, and $\lambda=0.9$ were taken as default settings \citep{AD3}. The whole DGHM-C Loss was finally defined as
\begin{equation}
	L_{d g h m-c}=\frac{1}{M N} \sum_{i=1}^{N} \beta_{i} L_{c e-d}\left(p^{(i)}, c^{(i)}\right),
	\label{eq:dghm-c}
\end{equation}
where $M$ was the number of gradient norm distributions.

Secondly, the regression loss functions for SRC detection are illustrated as follows. 

\textbf{Smooth L1 Loss} \citep{Fast-R-CNN} was applied in almost all SRC detectors. It was first proposed in Fast RCNN \citep{Fast-R-CNN} to replace L2 Loss used in SPPNet \citep{SPPNet} and R-CNN \citep{R-CNN}. Smooth L1 Loss alleviated sensitivity to outliers, so as to prevent gradient explosion in training. A four-dimensional vector $\vec{b}=<x, y, h, w>$ was required to be regressed for each predicted bounding box, where $x$ and $y$ were the normalized coordinates of the center of the bounding box, $h$ and $w$ represented the height and width of the bounding box, respectively. Similarly, $\vec{b_a}=<x_a,y_a,h_a,w_a>$ and  $\vec{b^\ast}={<x}^\ast,{\ y}^\ast,\ h^\ast,\ w^\ast>$ were introduced to encode an anchor box and the ground truth, respectively. Regression offsets could be calculated as follows:
\begin{equation}
	\left\{\begin{array}{ll}
		t_{x}=\left(x-x_{a}\right) / w_{a}, &t_{x}^{*}=\left(x^{*}-x_{a}\right) / w_{a} \\
		t_{y}=\left(y-y_{a}\right) / h_{a}, &t_{y}^{*}=\left(y^{*}-y_{a}\right) / w_{a} \\
		t_{h}=\log \left(h / h_{a}\right), &t_{h}^{*}=\log \left(h^{*} / h_{a}\right) \\
		t_{w}=\log \left(w / w_{a}\right), &t_{w}^{*}=\log \left(w^{*} / w_{a}\right)
	\end{array}\right. ,
\end{equation}
where $<t_x,t_y,t_h,t_w>$ denoted the offsets between the prediction and the anchor box, and $<t_x^\ast,t_y^\ast,t_h^\ast,t_w^\ast>$ denoted the offsets between the anchor box and the ground truth. Then, the Smooth L1 Loss was calculated as
\begin{equation}
	L_{\text {smoothL}_{1}}\left(\vec{t}, \vec{t}^{*}\right)=\sum_{j \in\{x, y, h, w\}} f\left(t_{j}-t_{j}^{*}\right),
\end{equation}
where $\vec{t}=<t_x,t_y,t_h,t_w>$, and ${\vec{t}}^\ast=<t_x^\ast,t_y^\ast,t_h^\ast,t_w^\ast>$, $ f(\cdot)$ was defined as
\begin{equation}
	f(x)=\left\{\begin{array}{cl}
		0.5 x^{2}, &\text { if }\lvert x\rvert<1 \\
		\lvert x\rvert-0.5, &\text { otherwise }
	\end{array}\right.
\end{equation}

Thirdly, other loss functions for SRC detection are illustrated as follows. 

Auxiliary embedding layers were also introduced in some SRC detection methods to learn discriminative features, such as similarity learning embeddings \citep{AD4, AD7} with pair loss \citep{loss-Pair} or triplet loss \citep{loss-Triplet}. We will introduce these two loss functions in this section.

\textbf{Pair Loss} \citep{loss-Pair}. In the task of SRC detection, Sun et al. use the Pair Loss to pull the anchors of the same category closer, while pulling away the anchors of different categories in the embedding space \citep{AD7}. Pair Loss was defined as
\begin{equation}
	\begin{array}{rl}
		L_{\text {pair}}\left(\sigma, \sigma^{\prime}, s\right)=& s\left\|\sigma-\sigma^{\prime}\right\|^{2}/2
		+(1-s) \\
		&\max \left(m-\left\|\sigma-\sigma^{\prime}\right\|^{2}, 0\right)/2,
	\end{array}
\label{eq:pair}
\end{equation}
where $\sigma$ and $\sigma^\prime$ represented the embeddings of two sampled anchors, $s\in\left\{0, 1\right\}$ denoted the closeness between them, $\left\| \cdot \right\|$ was the Euclidean distance metric, and $m$ was a constant of margin. It could be observed that as the loss function decreased, the distance between two samples of different categories should be greater than $m$, while samples of the same class were getting closer. It enabled the models to learn more discriminative features, which benefited the subsequent classification.

\textbf{Triplet Loss} \citep{loss-Triplet}. Triplet Loss was another popular loss function for similarity learning. Its purpose was similar to Pair Loss (Equation \ref{eq:pair}) with a different form calculated among three samples:
\begin{equation}
\begin{array}{r}
	L_{\text {triplet}}\left(\sigma^{a}, \sigma^{p}, \sigma^{n}\right)=\max (\left\|\sigma^{a}-\sigma^{p}\right\|^{2}
	\\-\left\|\sigma^{a}-\sigma^{n}\right\|^{2}+m, 0) ,
\end{array}
\label{eq:triplet}
\end{equation}
where $\sigma^a$ was a reference embedding, and $\sigma^p$ was a positive embedding of the same category with the reference while $\sigma^n$ represented a negative one of another category. After optimization, the distance between $\sigma^a$ and $\sigma^p$ would be less than that between $\sigma^a$ and $\sigma^n$ by a margin of $m$.

\subsubsection{Training strategy}
\textbf{Fully supervised learning.} Fully supervised learning was the most common training strategy in SRC detection \citep{AA2,AD1,AD3,AD4,AD5,AD6,AD7,B2302}. Models could achieve satisfactory performance with sufficient training data and high-quality annotations. However, DigestPath dataset suffered from a problem of incomplete labeling, which introduced noises during training stage. Besides, variation in color, shape, size, scale, and the overlaps of SRCs also brought great challenges to detection. Most of the existing studies implemented the SRC detection by fully supervised learning which relied entirely on accurate annotations of the training data. These solutions could be divided into two categories: methods based on modified loss functions and auxiliary modules. Among them, the loss functions were refined to make the models more robust. For example, Zhang et al. proposed the RGHMC Loss (Equation \ref{eq:rghmc}) with a label correction module which treated the revised ground-truth labels as the reference for calculating gradients \citep{AD1}. Lin et al. decoupled noisy samples from clean ones and devised the DGHM-C Loss (Equation \ref{eq:dghm-c}) to harmonize their gradient distributions respectively \citep{AD3}. In addition, some methods employed additional auxiliary modules to enable the model to learn discriminative features for enhancing the classification. For example, Wang et al. and Chen et al. designed a Classification Reinforcement Branch (CRB) to extract more comprehensive features containing cells and their surrounding context \citep{AD5, AD6}. Sun et al. introduced an embedding layer to perform similarity learning, which adopted Pair Loss (Equation \ref{eq:pair}) or Triplet Loss (Equation \ref{eq:triplet}) to narrow the intra-class distance and expand the inter-class one \citep{AD4, AD7}.

\textbf{Semi-supervised learning.} Semi-supervised learning aimed to handle the issue of lack of well-labeled data. However, among the current SRC detection methods, only Ying et al. applied the semi-supervised learning strategy \citep{AD2}. Specifically, a simple but efficient self-training framework was proposed to deal with the partial annotations in DigestPath dataset, which could be divided into three procedures. In the first step, an initial RetinaNet \citep{RetinaNet} model was trained with annotated labels. During the inference time, the pseudo bounding boxes were generated by the initial network, and then filtered by the novel Test Time Augmentation (TTA) and modified NMS algorithms to supplement the dataset with high-quality labels. With the new dataset, an iteratively retraining could be performed until there was no improvement of the detector. They achieved the 1st place in the SRC detection task of the Digestive-System Pathological Detection and Segmentation Challenge 2019, which demonstrated a superior potential of semi-supervised learning in this task.

\subsection{Image segmentation}
\label{subsec:segmentation}
\begin{figure*}[!thb]
	\centering
	\includegraphics[width=\textwidth]{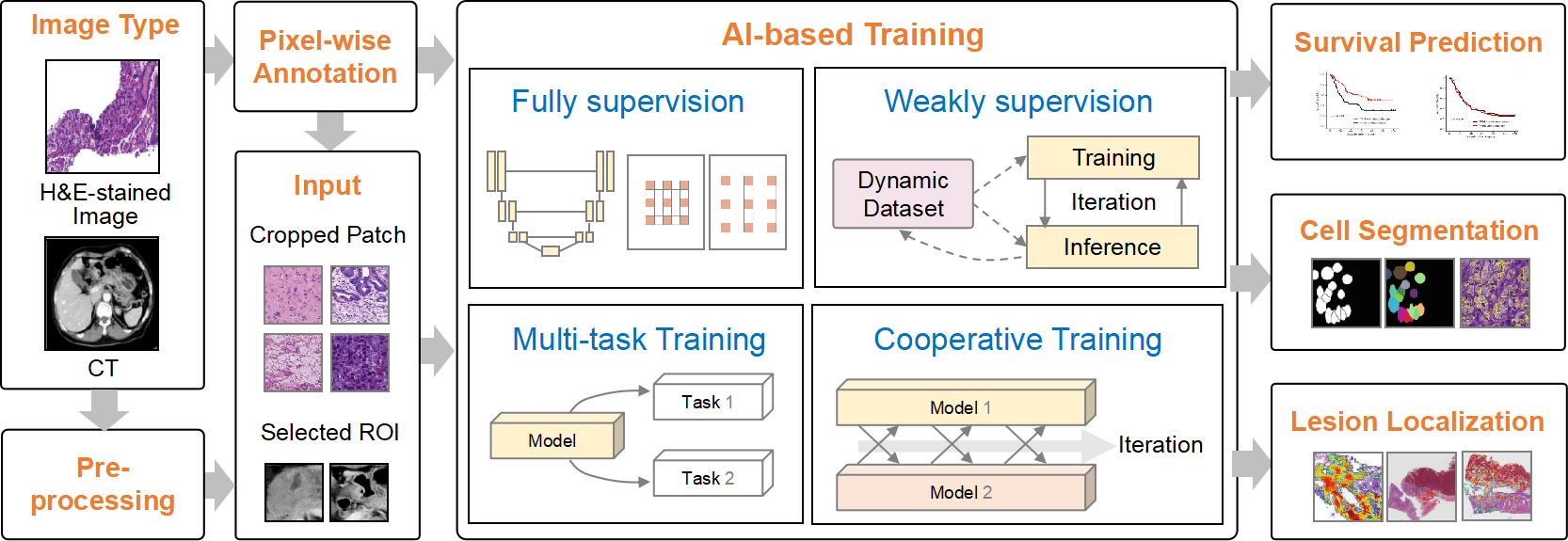}
	\caption{\label{fig:Seg}An overview of segmentation methods for SRC diagnosis.}
\end{figure*}

Segmentation clearly outlined the shape of the lesions. Unlike classification algorithms that mapped an input patch to a single category, segmentation algorithms mapped a patch to a diagnostic heatmap with the same size as the input patch. Therefore, segmentation algorithms could perform delicate and complex tasks such as cell segmentation. As shown in Fig. \ref{fig:Seg}, among the articles related to SRC diagnosis covered in this survey, the segmentation algorithms completed four training strategies, namely, full supervision, weak supervision, multi-task training, and cooperative training, to complete sub-tasks such as survival prediction, cell segmentation, and lesion localization. The overview of articles related to SRC segmentation is summarized in Table \ref{table:Seg}. The details of AI-based training strategies and task-related post-processing are presented next.

\onecolumn
\footnotesize
\begin{landscape}
\begin{longtable}{|m{1.9cm}<{\centering}|m{0.75cm}<{\centering}|m{1.5cm}<{\centering}|m{1.5cm}<{\centering}|m{2.5cm}<{\centering}|m{2.5cm}<{\centering}|m{2.75cm}<{\centering}|m{1.0cm}<{\centering}|m{4.0cm}<{\centering}|}
	\caption{\leftline{\label{table:Seg}Summary of automatic SRC diagnosis algorithms on the basis of segmentation.}}
 \\
	
	\hline 
	
	\textbf{Publication} & \textbf{Year} & \textbf{Modality} & \textbf{Target} & \textbf{Task} & \textbf{Data} & \textbf{Network} & \textbf{Loss} & \textbf{SRC diagnosis} \\ 
	
	\hline  
	
	\endfirsthead
	
	\hline 
	
		\textbf{Publication} & \textbf{Year} & \textbf{Modality} & \textbf{Target} & \textbf{Task} & \textbf{Data} & \textbf{Network} & \textbf{Loss} & \textbf{SRC diagnosis} \\ 
	
	\hline   
	
	\endhead
	
	\endfoot
	
	\endlastfoot

		\citep{AA1} & 2022 & CT & SRC carcinoma & SRC carcinoma diagnosis and chemotherapy response prediction & 855 images (the maximum cross sections of precontrast for each patient) & U-Net & CE-Loss & Segmentation encoder was assigned on only one single slice of a 3D matrix for each patient to diagnose SRC carcinoma. \\ \hline
		
		\citep{AS1} & 2019 & H\&E & SRC & SRC segmentation with a semi-supervised learning framework & 127 WSIs from 10 organs (at least 3 cropped regions for each WSI) & U-Net with ResNet-34 or DLA as the encoder & CE-Loss, 
		IOU-Loss & Multi-organ SRC segmentation with relatively small amount of annotation cost. \\ \hline
		
		\citep{AF6} & 2020 & H\&E & Gastric cancer & A two-class system to distinguish between benign and malignant & 6917 WSIs from 3 centers & DeepLab-v3 with ResNet-50 as the backbone & CE-Loss & SRC carcinoma was included in the malignant category and was easily overlooked when there were limited cancer cells. \\ \hline
		
		\citep{AF10} & 2021 & H\&E & Gastric cancer & Screening and localization of gastric cancer based on a multi-task CNN & 10,315 WSIs collected from 4 medical centers & DLA structure combined with classification and segmentation branches & BCE-Loss & SRC carcinoma was one type of the positive targets.  \\ \hline
		
		\citep{AF5} & 2022 & H\&E & Gastric cancer & Assessment of deep learning assistance for the pathological diagnosis & 110 WSIs and 16 board-certified pathologists  & DeepLab-v3 with ResNet-50 as the backbone & CE-Loss & SRC carcinoma was included in the malignant category and deep learning could help pathologists improve the diagnosis accuracy of scattered SRCs. \\ \hline
		
		\citep{AS2} & 2022 & H\&E & Gastric SRC carcinoma & Quantifying the cell morphology and predicting biological behavior & 607 WSIs from stomach or colorectum & A coarse U-Net to find approximate region at $10\times$ magnification, a fine U-Net to segment SRC at $40\times$ magnification & CE-Loss, 
		IOU-Loss & SRC segmentation was adopted to quantify the cell morphological characteristics and atypia so as to analyze the biological behavior of SRC carcinoma. \\ \hline

        \citep{B2350}&2022&H\&E&Lung Cancer&Prediction of ALK gene rearrangement in patients with non-small-cell lung cancer&300 WSIs from 208 patients&DenseNet-121 with up-sampling and skip connections&BCE-Loss&SRC carcinoma was considered in the histologic types of surgical specimen.\\\hline
        
        \citep{B2364}&2022&H\&E&Gastric cancer&Development and multi-institutional validation of an AI-based diagnostic system for gastric biopsy&984 patients for training and 2771 patients for validation&U-Net&Not mentioned&The performance of a poorly cohesive adenocarcinoma with SRCs was discussed.\\\hline
        
        \citep{B2319}&2023&H\&E&Colorectal cancer&Lymph node segmentation and metastasis detection&100 WSIs of lymph node regions from 14 patients for segmentation&U-Net&BCE-Loss, CE-Loss&In a validation cohort with mucinous and SRC histology, the performance was slightly worse because of limited examples.\\\hline
        
        \citep{B2346}&2023&H\&E&Colorectal cancer&Hybrid deep learning framework for tumor segmentation&DigestPath and GlaS \citep{GlaS} datasets&RGSB-UNet derived from U-Net, residual block, and  bottleneck Transformer&Class-wise Dice Loss&SRC carcinoma was regarded as one type of malignant lesions in this study. \\\hline

        \citep{J202510RGGC}&2024&H\&E&Gastric mucosa and intestine&Semantic segmentation of SRCs&308 high-resolution images, DigestPath and GlaS datasets&RGGC-UNet&Dice Loss&Specially designed a CNN architecture for SRCs. \\\hline

        \citep{J-seg-pro}&2025&H\&E&Bladder, prostate, seminal vesicle, lymph node&Deep visual proteomics analysis of SRC molecular signatures&More than 1000 SRCs annotated for training of instance segmentation after semantic segmentation&Fine-tuning a pre-trained model in BIAS\cite{BIAS} &Not mentioned&SRC segmentation is a key fundamental step in analyzing multi-organ SRC metastasis using spatial proteomics. \\\hline
\end{longtable}
\end{landscape}
\normalsize

\subsubsection{AI-based training strategies}
\textbf{Fully supervised training.} Segmentation put forward high requirements with accurate pixel-wise annotation. The fully supervised trained segmentation algorithms could both capture SRC-positive regions in CT images \citep{AA1} and localize lesion regions in H\&E stained images \citep{AS1, AF6, B2319, B2346}. U-Net \citep{U-Net}, DeepLab \citep{DeepLab} or their variants were usually used as the skeleton networks. U-Net was a U-shaped network with encoder-decoder structure originated from the task of cell segmentation. The encoder extracted high-level semantic information through continuous down-sampling, the decoder restored the image size by up-sampling, and the encoder transmitted spatial location information to the decoder through skip-connections. DeepLab used VGG, ResNet and other classic classification network structures as the basic backbone, expanded the receptive fields through atrous convolution, and finally improved the segmentation performance through conditional random field (CRF) \citep{CRF}. Then its variant DeepLab v3+ \citep{DeepLabV3Plus} removed the time-consuming CRF, and further improved the integration ability of spatial information through Atrous Spatial Pyramid Pooling (ASPP) \citep{AF10}. Segmentation algorithms often used CE-Loss (Equation \ref{equ:ce}) to constrain pixel-wise prediction. In addition, intersection over union loss (IOU-Loss) \citep{IOU-Loss} and Dice-Loss \citep{Dice-Loss} were popular for the positive region prediction. IOU-Loss $L_{iou}$ and Dice-Loss $L_{dice}$ were defined as 
\begin{equation}
	L_{iou}=1-\frac{\sum_{i}{y_ip_i}}{\sum_{i} y_i+\sum_{i}{p_i-\sum_{i}{y_ip_i}}} ,
\end{equation}
\begin{equation}
	L_{dice}=1-\frac{2\sum_{i}{y_ip_i}}{\sum_{i} y_i^2+\sum_{i} p_i^2} ,
\end{equation}
where $y_i$ and $p_i$ were the ground truth and predicted probability for the pixel $i$, respectively.

\textbf{Multi-task Training.} Classification was the global control of the input image patch, while segmentation was the local control of the details of that. Since classification and segmentation had complementary advantages, multi-task training could constrain model optimization from both global and local levels. For example, Yu et al. embedded two parallel branches after feature extraction in the backbone network to achieve classification and segmentation simultaneously \citep{AF10}. Multi-task learning could balance different requirements simultaneously.

\textbf{Weakly supervised training.} Although fully supervised segmentation algorithms could handle many segmentation requirements, there were often situations in which data annotations mismatched the task objectives in reality. For example, DigestPath dataset annotated SRCs with rectangular boxes that contained SRCs without outlining their contours. Therefore, Li et al. achieved SRC segmentation by weakly supervised learning \citep{AS1}. They approximated the contours of SRCs by generating inscribed ellipses within the bounding boxes. Although this operation inevitably led to labeling errors, the ellipses could be used as training data for the fully supervised segmentation methods described above. Given that there were a large number of unlabeled SRCs in the slides, the early-stage segmentation model was adopted to segment the unlabeled data, and some positive regions with the highest confidence probabilities were added to the training set for retraining. The training and inference alternated to achieve self-training, and then gradually improved the model accuracy.

\textbf{Cooperative training.} Different network structures often had different emphasis on feature extraction. Therefore, some algorithms fused the inference output of different models to improve the final diagnostic performance. For example, Li et al. replaced the encoder of U-Net with ResNet and DLA \citep{DLA}, respectively, resulting in two different networks \citep{AS1}. Then, the self-training strategy mentioned above was adopted for the two networks. In the iterative process, the inference results of the other network were used to strengthen the training, so as to realize the information exchange of different networks. The two models converged to be consistent in performance while improving accuracy. Such a cooperative training strategy could avoid a single model from getting stuck in local optima.

\subsubsection{Task-related post-processing}
\textbf{Survival prediction.} The segmentation algorithms directly presented the SRC positive regions, which was very convenient for the evaluation of SRC aggregation degree, lesion area and other metrics. Therefore, segmentation played an important auxiliary role in the formulation of clinical diagnosis and treatment strategies. For example, Li et al. extracted SRC carcinoma features through segmentation models to identify patients who could benefit from postoperative chemotherapy based on preoperative contrast-enhanced CT \citep{AA1}.

\textbf{Cell segmentation.} SRCs could either aggregate in clusters or distribute in isolation. Segmentation methods that took the lesion area as a whole tend to miss isolated SRCs. In addition, when suggesting the existence of SRCs, providing quantitative information such as the degree of cell aggregation was more interpretable and convincing clinically. Therefore, cell segmentation for SRCs was an efficient and intuitive basis for quantitative analysis. The challenge of cell segmentation was that cell-level annotation was time-consuming and labor-intensive. Li et al. utilized weakly supervised learning and cooperative training to train a cell segmentation model on the basis of bounding box annotations \citep{AS1}. Then, Da et al. adopted the segmented results to measure the inherent properties of SRCs including the cross-sectional area of cell plasma and nuclear, the cell ellipticity, as well as the nuclear / cytoplasmic ratio \citep{AS2}. Kabatnik et al. implemented Deep Visual Proteomics (DVP), integrating AI-guided cell segmentation with laser microdissection and ultra-high sensitivity mass spectrometry to characterize the proteomic profiles of SRCC across multiple organs in a single patient, revealing dysregulated DNA damage response pathways and immunogenic tumor microenvironment that suggest potential responsiveness to PD-1 blockade therapy \citep{J-seg-pro}.

\textbf{Lesion segmentation.} Similar to lesion classification, lesion segmentation combined patch-level results into a diagnostic heatmap corresponding to the slide by splicing. The difference was that the segmentation was a pixel-wise classification, that is, the size of the diagnostic heatmap was the same as the original slide. Due to the mining of details by the segmentation algorithms, it was possible to focus on small lesions. Song et al. developed a gastric cancer lesion segmentation system based on DeepLab v3+ \citep{AF6}. Ba et al. further demonstrated that the system assistance indeed improved pathologists’ accuracy in gastric cancer diagnosis \citep{AF5}. However, the components in H\&E stained slides were complex. Many factors led to the inevitable false positive noise or false negative holes in pixel-by-pixel analysis. Therefore, Yu et al. embedded both classification and segmentation branches in the backbone network. The classification branch was used to control global features, and the segmentation branch was used to refine local features, so as to obtain diagnostic heatmap results that were more in line with the actual diagnosis experience of pathologists \citep{AF10}. 

\subsection{Image foundation models}
\label{sec:imageFM}

\Rone{The field of computational histopathology has witnessed transformative advances through self-supervised foundation models which demonstrate remarkable capability in extracting generalizable features from unannotated WSI datasets at scale. Representative models such as UNI} \citep{UNI} \Rone{employed DINOv2} \citep{dinov2} \Rone{self-supervision frameworks trained on Mass-100K, a landmark dataset containing over 100 million tissue patches derived from 100,426 diagnostic WSIs spanning 20 major tissue types. Complementing this approach, the BEPH model} \citep{BEPH} \Rone{leveraged masked image modeling pre-training on 11.77 million histopathological patches derived from 11,760 WSIs across 32 TCGA cancer types. Despite utilizing a dataset an order of magnitude smaller than ImageNet, BEPH demonstrated competitive performance across diverse downstream tasks, achieving large improvements over traditional CNNs and weakly supervised models in patch-level classification. This extensive coverage inherent in models like UNI and BEPH captured diverse morphological patterns, including potentially rare ones like SRCs. Benchmark evaluations across numerous clinical tasks, encompassing cancer subtyping, metastasis detection, and biomarker screening, confirmed that foundation models like UNI and BEPH established new SOTA performance metrics, particularly excelling in label efficiency, resolution robustness, and rare cancer classification compared to conventional encoders like ResNet. In addition to publicly available datasets, the industry is increasingly focusing on the true value of these foundation models in clinical practice and emphasizing the evaluation effects of various downstream tasks on real data from different cohorts, proving that these foundation models have efficient image representation capabilities. Crucially, rigorous clinical benchmarking from multiple institutions indicated that performance gains for both disease detection and biomarker prediction exhibited diminishing returns with increasing model scale and pre-training dataset size, challenging direct applicability of scaling laws observed in other domains to computational pathology. Instead, the composition and tissue-specific relevance of pre-training data emerged as critical determinants of downstream performance} \citep{benchmark}.

\Rone{Domain-specific foundation models address subspecialty challenges through targeted strategy innovations. In gastrointestinal pathology, Digepath} \citep{DigePath} \Rone{implemented a dual-phase training paradigm: initial multi-scale self-supervised learning across 210,043 gastrointestinal WSIs captured resolution-dependent histological features, followed by dynamic RoI mining that iteratively optimized diagnostic discriminators. This specialized approach achieved superior performance on 33 of 34 gastrointestinal tasks, demonstrating the efficacy in identification of microscopic SRC foci within screening contexts. This focus on domain-relevant data curation aligned with benchmark findings emphasizing tissue-specific data importance} \citep{benchmark}.

\Rone{Recent architectural breakthroughs further enhanced computational efficiency without compromising diagnostic accuracy. Lightweight transformer architectures such as PathDino} \citep{PathDino} \Rone{incorporated HistoRotate augmentation to establish 360° rotation invariance, effectively mitigating overfitting while maintaining performance parity with larger models in metastasis detection tasks. 
These insights delineate two complementary paradigms for SRC diagnosis: generalist foundation models such as UNI, BEPH, and PathDino, leverage pan-cancer morphological diversity to construct versatile feature extractors, while specialized architectures like Digepath employ domain-optimized data curation and iterative refinement to capture diagnostic subtleties. These models reduce dependence on scarce expert annotations through effective self-supervised learning on large unlabeled datasets. Their synergistic integration establishes robust computational backbones for SRC diagnostic pipelines, simultaneously reducing dependence on scarce expert annotations and enhancing cross-institutional generalizability.}

\subsection{‌Omics-based algorithms}
\Rone{Omics technologies establish a multi-dimensional analytical framework for the molecular diagnosis of SRCs, extending conventional histopathology. For instance, genomics comprehensively characterizes genetic alterations to delineate molecular subtypes of SRC; transcriptomics deciphers disease-associated gene expression profiles with therapeutic implications; proteomics accurately defines tumor-specific proteomic signatures; microbiomics further elucidates correlations between intra-tumoral microbiota and SRC pathogenesis within the tumor microenvironment; metabolomics dynamically captures metabolic reprogramming events during disease progression} \citep{J0-CLDN,J0-CXCL,J0-pancreas,J0-mc,O-RNA-TME,J0-discus-micro}. 

\Rone{‌In traditional research methodologies without AI, investigations into the prognosis of SRC typically employed classical biostatistical and molecular biology approaches. Taking the study of DNA MMR status impact on SRC prognosis as an example, researchers systematically compared clinicopathological characteristics among patients with different MMR statuses using Chi-square tests; then, survival outcomes were analyzed via Kaplan-Meier curves coupled with log-rank tests, and independent prognostic factors were identified by constructing Cox proportional hazards regression models} \citep{O-RNA-NC}. \Rone{Although these methods do not involve advanced AI technologies such as deep learning, rigorous statistical inference and bioinformatic analysis effectively illuminate key biological characteristics, notably tumor heterogeneity.}

\Rone{Recent advances in omics research have witnessed a strategic resurgence of classical machine learning algorithms, driven by their unique suitability for high-dimensional sequence data analysis, as summarized in Table} \ref{table:omics}. \Rone{Unlike computationally intensive deep learning approaches, these methods offer three distinct advantages in omics applications: (1) interpretability of decision pathways critical for biomedical validation, (2) stable performance with limited training samples, and (3) efficient feature selection capabilities for high-dimensional datasets. This computational parsimony makes them valuable for research settings where both analytical precision and resource constraints need to be balanced. For example, Fan et al. employed ‌three random forest classifiers to analyze proteomic data from gastric cancer tissues and cell lines} \citep{O-Pro-RF}. \Rone{Zhao et al. employed ‌Bayesian additive regression trees (BART)‌ to develop a risk prediction model integrating clinicopathologic, immune, microbial, and genomic data (75 variables) for stratifying survival outcomes in stage II–III colorectal cancer patients, achieving prognostic differentiation and external validation via TCGA dataset} \citep{O-BART}. \Rone{Yu et al. employed ‌random survival forests and ‌least absolute shrinkage and selection operator (LASSO) regression to identify key prognostic factors such as age, tumor size, stage, site, surgery, and chemotherapy, for pancreatic SRC carcinoma} \citep{O-pancr}. \Rone{Koppad et al. utilized ‌six machine learning classifiers (AdaBoost, ExtraTrees, logistic regression, naïve Bayes, random forest, and XGBoost)‌ to identify 34 diagnostic biomarker genes for colorectal cancer through transcriptomic analysis of GEO datasets, with random forest achieving optimal performance in differential gene expression classification} \citep{O-biology}. \Rone{Lin et al. analyzed ‌eight feature ranking algorithms (AdaBoost, CatBoost, ExtraTrees, LASSO, LightGBM, RF, Ridge, XGBoost)‌ combined with ‌incremental feature selection and ‌random forest classification to identify key miRNA-mRNA biomarkers differentiating ALK-positive from ALK-negative lung adenocarcinoma} \citep{O-mRNA}. \Rone{Ellrott et al. applied ‌five machine learning approaches (AKLIMATE, CloudForest, SK Grid, JADBio, and subSCOPE)‌ to develop 737 containerized predictive models for classifying 8,791 TCGA tumor samples across 26 cancer types into 106 molecular subtypes, achieving robust performance through cross-validation while prioritizing compact gene-centric feature sets to facilitate clinical translation of multi-omics data} \citep{O-cancer-cell}.

\onecolumn
\footnotesize
\begin{landscape}
\begin{longtable}{|m{1.0cm}<{\centering}|m{0.75cm}<{\centering}|m{1.8cm}<{\centering}|m{1.5cm}<{\centering}|m{2.5cm}<{\centering}|m{2.5cm}<{\centering}|m{1.45cm}<{\centering}|m{3.0cm}<{\centering}|m{4.0cm}<{\centering}|}
	\caption{\leftline{\label{table:omics}Summary of automatic SRC diagnosis algorithms on the basis of omics data.}}
 \\
	
	\hline 
	
	\textbf{Publi-cation} & \textbf{Year} & \textbf{Modality} & \textbf{Target} & \textbf{Task} & \textbf{Data} & \textbf{Method type} & \textbf{Method} & \textbf{Key findings} \\ 
	
	\hline  
	
	\endfirsthead
	
	\hline 
	
		\textbf{Publi-cation} & \textbf{Year} & \textbf{Modality} & \textbf{Target} & \textbf{Task} & \textbf{Data} & \textbf{Method type} & \textbf{Method} & \textbf{Key findings} \\ 
	
	\hline   
	
	\endhead
	
	\endfoot
	
	\endlastfoot

		\citep{O-Pro-RF} & 2019 & Prote-omics & Gastric carcinoma & Comparative analysis of proteomic profiles between gastric cancer tissues and cell lines & 14 cases of SRC carcinoma and 34 of adenocarcinoma (17 poorly and 17 well-moderately differentiated), with 6,639 proteins quantified & Machine learning & Three random forest classifiers & Classification accuracy for SRC carcinoma and adenocarcinoma was acceptable across the dataset. \\ \hline
		
		\citep{O-biology} & 2022 & Transcript-omics & Colorectal cancer & Identification of diagnostic biomarker genes for colorectal cancer & Three gene expression datasets (GSE44861, GSE20916, GSE113513) from the GEO database & Machine learning & Logistic regression, naïve Bayes, random forest, ExtraTrees, AdaBoost, and XGBoost & 34 genes were identified by the random forest algorithm as potential diagnostic markers for colorectal cancer. \\ \hline
		
		\citep{O-BART} & 2023 & Clinico-pathologic, immune, microbial, and genomic data & Colorectal cancer & Survival prediction for colorectal cancer patients & 815 stage II–III colorectal adenocarcinoma patients from two U.S.-wide prospective cohorts and 106 from TCGA for external validation & Machine learning & Bayesian additive regression trees (BART) & The BART model achieved superior performance compared to other machine learning methods.
 \\ \hline
		
		\citep{O-pancr} & 2025 & Clinico-pathological, treatment, and demographic data & Pancreatic SRC carcinoma & Post-chemotherapy survival analysis in patients with pancreatic SRC carcinoma & 708 patients from the SEER database & Machine learning & Random survival forests and least absolute shrinkage and selection operator (LASSO) regression & Six key prognostic factors were identified in patients with pancreatic SRC carcinoma.
  \\ \hline
		
		\citep{O-mRNA} & 2025 & Transcript-omics & Lung adenocarcinoma & Identification of unique molecular characteristics in ALK-positive lung adenocarcinoma & Expression profiles of 77 lung adenocarcinoma patients from the GEO database (GSE128311) & Machine learning & Eight feature ranking algorithms (AdaBoost, CatBoost, ExtraTrees, LASSO, LightGBM, RF, Ridge, XGBoost) & Key differentially expressed genes and miRNAs were identified as potential biomarkers.
 \\ \hline
		
		\citep{O-cancer-cell} & 2025 & Multi-omics (mutation, copy number, mRNA, DNA methylation, and miRNA) & 26 cancer types & TCGA molecular subtype classification of tumors & 8,791 TCGA tumor samples from 26 cancer types, covering 106 molecular subtypes & Machine learning & 737 predictive models developed using five machine learning methods (AKLIMATE, CloudForest, JADBio, SK Grid, and subSCOPE) & The models showed robust performance with compact feature sets and enabled clinical subtype classification of non-TCGA tumor samples.
        \\\hline
        
        \citep{O-life} & 2023 & Transcript-omics & Gastric cancer & Differentially expressed gene analysis & Single-cell RNA sequencing data of 46,883 gastric cancer cells & Un-supervised clustering & Graph-based clustering & Differentially expressed genes were identified across distinct cell types.
 \\ \hline

        \citep{O-cluster} & 2025 & Immunomics (tumour-infiltrating immune cell profiles) & Gastric cancer & Prognostic evaluation in gastric cancer patients based on immune variables & 371 patients from Henan Cancer Hospital & Un-supervised clustering & CLARA & The model demonstrated effectiveness in patient risk stratification and prognosis prediction.
\\\hline
        
        \citep{Flexynesis} & 2024 & Multi-omics & Precision oncology & Development of a flexible and adaptable framework integrating bulk multi-omics data for various predictive tasks. & Extensive benchmark datasets including CCLE, GDSC2, LGG, GBM, and TCGA & Deep learning & Integrated multiple models including MLPs, variational autoencoders (VAEs), graph neural networks (GNNs), and cross-modality encoders & The framework proved effective in regression, classification, survival analysis, and unsupervised tasks, featuring automated hyperparameter tuning, multi-task modeling with missing label tolerance, and domain adaptation.
\\\hline
        
        \citep{O-DOMSCNet} & 2025 & Multi-omics (Exon, mRNA, miRNA expression, and DNA methylation) & Stomach cancer & Classification of stomach cancer and normal tissue & Four multi-layer omics datasets from TCGA-STAD (Exon, mRNA, miRNA, DNA methylation); eight external datasets from NCBI GEO and TCGA-LIHC for validation & Deep learning & DOMSCNet (based on recurrent neural networks) & DOMSCNet outperformed existing models across all multi-layer omics datasets.
 \\\hline

        \citep{T-BulkFormer} & 2025 & Transcript-omics & General disease modeling & Bulk transcriptome modeling across diverse downstream tasks & Over 500{,}000 human bulk RNA-seq profiles, covering about 20{,}000 protein-coding genes & Foun-dation model & Hybrid architecture (combining GNNs and performer modules) & BulkFormer achieved strong performance in six downstream tasks: transcriptome imputation, disease annotation, prognosis modeling, drug response prediction, compound perturbation simulation, and gene essentiality scoring.
 \\\hline

        \citep{T-Omics-LLM} & 2025 & Clinical and multi-omics data & Various cancer types & Multi-omics clustering for cancer subtyping & Six cancer datasets on three omics levels & Foun-dation model, deep learning & BERT (for clinical feature extraction), autoencoder (for feature integration) & The integration of clinical features significantly improved clustering performance, surpassing current approaches in cancer subtyping.
 \\\hline
\end{longtable}
\end{landscape}
\normalsize

‌\Rone{Unsupervised clustering algorithms demonstrate unique advantages in multi-omics analysis by eliminating the dependency on annotated datasets while revealing intrinsic biological patterns.‌ As evidenced by Zhou et al., single-cell RNA sequencing data from 46,883 gastric cancer cells were analyzed through clustering, successfully identifying differentially expressed genes of different cell types without requiring pre-labeled training data} \citep{O-life}. \Rone{Similarly, Hu et al. employed CLARA, a robust unsupervised algorithm, to stratify 371 gastric cancer patients into three prognostic subgroups based solely on tumor-infiltrating immune cell profiles} \citep{O-cluster}. \Rone{These studies highlight how unsupervised methods circumvent the limitations of supervised approaches that demand extensive clinical annotations, while enabling discovery of novel biomarkers.

Moreover, multi-omics data can also be effectively interpreted through deep learning algorithms. DOMSCNet} \citep{O-DOMSCNet}, \Rone{a deep recurrent neural network-based model, demonstrated robust performance in stomach cancer classification by integrating Exon expression, mRNA expression, miRNA expression, and DNA methylation data.} \Rone{Concurrently, Flexynesis}\citep{Flexynesis} \Rone{addressed broader challenges in bulk multi-omics integration through a modular framework that combines diverse architectures, including MLPs, variational autoencoders (VAEs), graph neural networks (GNNs), and cross-modality encoders, to support regression, classification, survival analysis, and unsupervised tasks. Flexynesis distinguished itself with features like automated hyperparameter tuning, multi-task modeling with missing label tolerance, and fine-tuning capabilities for domain adaptation.}

\Rone{Foundation models have demonstrated exceptional performance in multi-omics analyses, mirroring their remarkable achievements in natural language processing and computer vision domains. This consistent excellence stems from their fundamental capability to identify intricate patterns across high-dimensional data spaces. BulkFormer} \citep{T-BulkFormer} \Rone{is a 150M-parameter foundation model for bulk transcriptome analysis, pretrained on more than 500,000 human bulk RNA-seq profiles covering about 20,000 protein-coding genes. Its hybrid architecture combines GNNs (modeling gene-gene interactions from biological graphs) and performer modules (capturing global expression dependencies), enhanced by rotary expression encoding to preserve expression magnitude and continuity. BulkFormer consistently performs well in all six downstream tasks: transcriptome imputation, disease annotation, prognosis modeling, drug response prediction, compound perturbation simulation, and gene essentiality scoring. Ye et al. developed an innovative cancer subtyping framework that synergistically combined BERT-processed clinical narratives from pathology reports with multi-omics molecular profiles through an attention-guided autoencoder architecture, subsequently employing singular value decomposition and spectral clustering to achieve superior subtype discrimination across six major cancer types} \citep{T-Omics-LLM}.

\subsection{‌Text-based algorithms}
\Rone{The integration of text-based AI algorithms for SRC diagnosis has evolved with the advent of LLMs. While general-purpose LLMs like ChatGPT-4.0 and Gemini Advanced demonstrate robust natural language processing capabilities} \citep{healthLLM}, \Rone{their application to SRC-specific tasks requires domain-specific adaptations. These models excel at parsing histopathological descriptors (e.g., ``cytoplasmic mucin vacuoles" or ``eccentrically displaced nuclei") from electronic health records, but face limitations in precision due to hallucination and inadequate contextual understanding of rare entities like SRCs. Specialized text algorithms now employ fine-tuned biomedical language architectures such as BioClinicalBERT to optimize feature extraction from pathology reports. For instance, Jain et al. highlighted LLMs' ability to identify SRC terminology in differential diagnoses} \citep{SRC-LLM}. \Rone{Despite the growing capability of LLMs in text comprehension, pathological reports still require human diagnosis and drafting, which remains heavily reliant on pathologists' expertise without achieving true intelligent automation. Therefore, integrating image data into AI-assisted SRC diagnosis is imperative for future advancements.}

\section{Multi-modal algorithms}
\label{sec:multimodal}
\Rone{Diagnostic intelligence for SRC extends beyond unimodal approaches, leveraging AI algorithms to uncover nuanced and synergistic information across complementary data modalities, thereby achieving higher diagnostic accuracy aligned with real-world clinical practice. Given pathology's status as the diagnostic gold standard, our survey deliberately focuses on multi-modal fusion strategies which particularly integrate histopathological images with textual reports and multi-omics data. While multi-modal models typically demand substantial data across modalities and often designed for broad disease coverage, they nonetheless prove effective for characterizing SRC.}

\subsection{Collaboration between images and text}
\Rone{Aligning histopathology image representations with textual descriptions constitutes a relatively prevalent multi-modal fusion paradigm in computational pathology. The text embedding strategy employed by CHIEF} \citep{M2-CHIEF} \Rone{demonstrated notable advantages in computational efficiency and implementation simplicity, particularly through its direct extraction of anatomical site descriptors followed by CLIP} \cite{CLIP}-based \Rone{feature encoding. This approach effectively established organ-level discriminative capabilities by aligning visual histopathological patterns with standardized anatomical text prompts (e.g., ``This is a histopathological image of [organ]"), which significantly reduced the need for manual annotation while maintaining robust performance in WSI analysis. However, the algorithm exhibited inherent limitations due to its static text encoding paradigm where identical vector representations were generated for each anatomical category regardless of pathological subtypes, consequently constraining its diagnostic granularity for cancer subclassification where subtle textual variations in histopathological reports could provide critical discriminative signals. This trade-off between operational simplicity and subtype-specific adaptability reflected the fundamental design compromise in current multi-modal foundation models for computational pathology. In contrast, ‌ConcepPath} \citep{M2-ConcepPath} \Rone{addressed this granularity limitation by dynamically inducing disease-specific textual concepts from medical literature through GPT-4, then aligning them with histopathology patches via a CLIP-based model. Unlike CHIEF’s static organ-level prompts, ConcepPath’s hierarchical aggregation leveraged both expert-derived concepts and learnable data-driven concepts, enabling subtype-discriminative analysis through a two-stage fusion of patch-level concept scores and slide-level class prompts. Unlike the focus on characterizing local image patches, Prov-GigaPath} \citep{M2-Prov-GigaPath} \Rone{and PRISM} \citep{M2-PRISM} \Rone{emphasized the necessity of integrating global WSI information. TITAN} \citep{M2-TITAN} \Rone{achieved general-purpose representation learning for WSIs through self-supervised learning and vision-language alignment, with its cross-modal retrieval and report generation capabilities surpassing PRISM.

As a representative work in pathology multi-modal learning, CONCH} \citep{M2-CONCH} \Rone{employed a dual-stream Transformer architecture. The visual branch leveraged a hierarchical ViT framework to extract morphological features from histopathological WSI, while the textual branch utilized a pre-trained language model to encode clinical descriptive narratives. Cross-modal alignment was jointly optimized through a contrastive loss function and masked language modeling objectives. The resultant cross-modal representations effectively supported medical imaging analysis tasks including tissue grading and pathological report generation. Similarly, MUSK} \citep{M2-MUSK} \Rone{is another popular multi-modal foundational model; it achieved cross-modal deep alignment via a dual-stage pre-training architecture, involving unified masked modeling based on 50 million pathological images and 1 billion text tokens, and contrastive learning with 1 million image-text pairs.

Building upon these large-scale foundation models, more refined approaches for image and text representation have been developed. For instance, ALPaCA} \citep{M2-ALPaCA} \Rone{utilized GPT to structure pathology reports and generate both closed-ended and open-ended question-answer pairs. Subsequently, it utilized the Llama model for text representation while leveraging CONCH for pathological image feature extraction, ultimately achieving cross-modal representation alignment. While preserving the foundational visual representation capabilities of the UNI model, OmniPath} \citep{M2-OmniPath} \Rone{achieved enhanced parsing of histopathological image details and multi-scale features through two strategies: First, the Mixed Task-Guided Feature Enhancement module incorporated diagnostic prior knowledge into the feature encoding process by jointly optimizing auxiliary localization and segmentation tasks. Second, the Prompt-Guided Feature Completion system dynamically parsed textual descriptors from pathology reports, subsequently amplifying feature expression of tissue substructures through selective reinforcement. This synergistic dual-module architecture transcended the static representation constraints of conventional visual encoders, establishing a context-aware alignment mechanism grounded in clinical diagnostic semantics. Similarly, MR-PLIP} \citep{M2-‌MR-PLIP} \Rone{addressed the latent space representation discrepancies in histopathological images from the UNI model across varying scales through its ‌multi-resolution visual-textual alignment mechanism and ‌hierarchical feature consistency constraints.}

\subsection{Collaboration between images and omics}
\Rone{Histopathological images encapsulate morphological information at the megapixel scale, whereas omics data exist as low-dimensional feature vectors, leading a fundamental heterogeneity that impedes unified representation. To bridge this gap, OmiCLIP} \citep{M1-OmiCLIP} \Rone{innovatively adapted the CLIP architecture to biomedicine through its core dual-pathway encoders: The visual branch employed ViT to process H\&E-stained tissue sections, while the omics branch converted high-dimensional transcriptomic profiles into gene token sequences. By enforcing embedding space alignment via contrastive learning objectives on image-transcriptome pairs, this framework established latent semantic mappings between tissue morphology and molecular expression, ultimately enabling cross-modal retrieval. In contrast, MISO} \citep{M1-MISO} \Rone{adopted a divide-and-conquer strategy: It first trained modality-specific MLPs for individual modalities to generate structure-preserving low-dimensional embeddings. Subsequently, it computed outer products of paired modality embeddings to explicitly construct interaction tensors capturing nonlinear cross-modal relationships. Finally, it concatenated modality-specific features with interaction features into a comprehensive embedding. This architecture supported arbitrary modality combinations and enhanced robustness through manual filtering of low-quality features.}

\subsection{Collaboration of images, text, and omics}
\Rone{Joint representation learning across histopathology images, text, and omics data is emerging as a popular trend. ‌GECKO‌} \citep{M3-GECKO} \Rone{aligned WSIs with ‌pathology concept priors (derived from LLM-generated textual descriptions) via a ‌dual-branch MIL network (deep-encoding branch and concept-encoding branch), while seamlessly integrating auxiliary modalities like ‌transcriptomics data. ‌ModalTune‌} \citep{M3-ModalTune} \Rone{integrated ‌WSIs with multi-omics data  like transcriptomics via ‌Modal Adapters, while unifying multi-task learning through ‌LLM-generated text embeddings, specifically addressing catastrophic forgetting during fine-tuning and underutilization of shared information across tasks and modalities. Song et al. extracted multi-modal embedding features through zero-shot foundation models (UNI2 for histopathology images, BioMistral for pathology report text embeddings, and BulkRNABert for RNA-seq analysis), employing a linear Cox proportional hazards model with late fusion strategy to achieve cross-modal complementary cancer survival prediction in TCGA data, while optimizing text modality performance via Llama-generated pathology report summaries} \citep{M3-PCA}. ‌\Rone{spEMO‌} \citep{M3-spEMO} \Rone{integrated embeddings from pathology foundation models (e.g., GPFM, UNI) and LLMs with spatial multi-omics data (gene expression and protein profiles). It employed a ‌dual-framework design combining zero-shot learning and fine-tuning to achieve tasks including spatial domain identification, disease prediction, and cross-modal alignment.}

\subsection{Other collaborations}
\Rone{The Google Research and Google DeepMind team developed the ‌Med-Gemini‌} \citep{MM-Gemini} \Rone{series of medical models based on Gemini's multi-modal foundation. By fine-tuning the models with ‌2D or 3D histopathology, radiology images, ophthalmology, dermatology, and genomic data, they optimized multi-modal understanding for clinical applications. The study employed ‌modality-specific vision encoders (2D, 3D, and genomic)‌ and ‌instruction-tuning strategies, achieving SOTA performance on ‌17 out of 20 medical visual question answering tasks. }

\Rone{HistoXGAN‌} \citep{MM-GAN} \Rone{is a custom GAN that integrated ‌self-supervised learning pathology feature extractors with an ‌enhanced StyleGAN2 generator to achieve ‌precise pan-cancer histology reconstruction from pathologic, genomic, and radiographic latent features. The model employed ‌modality-specific vision encoders and an ‌L1 feature-consistency loss optimization strategy, validating across 29 cancer subtypes while preserving critical biologic traits such as tumor grade and histologic subtype. Notably, it pioneered the generation of ``virtual biopsy" tissue sections directly from MRI radiomic features.}

\section{Discussion and future outlook}
\label{sec:discussion}
\subsection{SRC datasets for learning}
The fitting of models based on deep learning was driven by big data, which put forward high requirements for the quantity and quality of data. There were a few studies on SRC automatic diagnosis in CT, MRI, and endoscopy images, and they were conducted on private datasets \citep{AA1, AA2,AA3,AA4, B2301, B2363}. More studies focused on accurate identification of SRCs in histopathological images, including the public datasets TCGA and DigestPath, and some private datasets. Specifically, in the task of gastrointestinal tract malignancy discrimination, although the number of patients in the datasets was large due to high morbidity, SRC carcinoma was only a special type of positive samples with few cases \citep{AF1, AF2, AF3, AF7, AF9, AF10, AW1, AF6}. The imbalance of samples in the training set often led to the inevitable omission of valuable SRCs in inference \citep{AF1, AF2, AF3, AW1, AF6}. In addition, the training set of the DigestPath dataset dedicated to SRC detection was composed of images from only 20 positive WSIs. Compared with natural images, the data scale of SRCs was very limited. For example, the emergence of ImageNet \citep{ImageNet}, a dataset with one million natural images, promoted the birth of excellent classification networks such as AlexNet, ResNet, VGG, and GoogLeNet. Recently, self-supervised learning has attracted much attention. Among them, the typical algorithm MoCo \citep{MoCo} captured the effective features of images through pretext tasks in Instagram-1B \citep{Dataset-Moco}, a dataset containing about one billion images. However, the privacy of medical data and the time-consuming of accurate labeling inevitably limited the size of the SRC datasets, and thus inhibiting the performance promotion of fully supervised learning \citep{S13}. To increase the robustness of the models as much as possible, when classical networks such as VGG and Inception were used to automatically diagnose SRCs, the parameters pretrained on the natural image datasets were often loaded before training to avoid the model overfitting or falling into local optimal solutions. In addition, to increase the generalization of the models, data augmentation was often embedded in the pre-processing process. Common data augmentation methods used in SRC pre-processing included random flipping, rotation, and color fluctuation. These methods improved the accuracy of the models from the perspective of technology, but they did not increase the number of real cases in essence.

In the future, the size of the datasets will not be limited to absolute numbers, but focus on the size of effective high-quality data. Current data augmentation methods did not enable the models to see diverse real-world cases, and the generalization performance remained limited during inference. Therefore, new and clinically challenging datasets should be constructed, in which the following four points may be paid special attention. 
\begin{itemize}
	\item To evaluate the generalization and robustness of the models in practical applications, the data are required to be derived from different medical centers and cover multiple scenes such as biopsy and surgical specimens.
	\item Since SRCs are lesions that can develop in various tissues and organs, the cases in the datasets should break through the common limitations of the gastrointestinal tract, and clearly indicate the primary and metastases.
	\item Typical SRC distributions may be involved, including aggregated and isolated cases.
	\item Multi-modal data are expected to be matched, such as age, gender, genes, and survival time, to facilitate further clinical studies.
\end{itemize}

\Rone{The severe scarcity of histopathologically confirmed SRC carcinoma cases presents a fundamental constraint for training data-hungry large deep learning models. While GANs and their variants} \citep{GAN} \Rone{offer potential solutions through style normalization and image synthesis, demonstrably enhancing model performance, their uncontrolled generation mechanisms frequently produce biologically implausible features. This limitation has precipitated significant data trust crises and ethical concerns regarding clinical applicability} \citep{S13, S14}. \Rone{Advancements in diffusion models will enable more controlled synthesis of pathologically credible SRC images. Future efforts should prioritize two biologically grounded strategies to enhance generative fidelity:}
\begin{itemize}
    \item \Rone{Quantitative modeling of morphometric relationships: Precise computational characterization of SRC-specific nuclear-vacuolar spatial configurations (e.g., nuclear displacement indices, vacuole-to-cytoplasm ratio) must guide image generation. This will permit granular synthesis of stochastic morphological variations beyond clinically documented presentations while preserving pathobiological validity.}
    \item \Rone{Anatomically conditioned synthesis: Given SRC’s propensity for diffuse infiltration across diverse organs (e.g., stomach, breast, peritoneum), generators should be conditioned on organ-specific clinical priors—including radiological localization, omics profiles, and histopathology reports. Such conditioning ensures anatomically faithful rendering of critical features like nuclear-cytoplasmic ratios and tumor cellularity, particularly for the diagnostically challenging solitary-diffuse growth patterns characteristic of metastatic SRC.}
\end{itemize}


\subsection{Selection of training strategy}
Fully supervised learning was considered robust and effective in natural image processing with sufficient training data. Despite the relatively limited amount of data in medical images compared to natural images, the models could still capture the typical morphological and structural features in the training data. When the input images had explicit expert annotations, fully supervised learning allowed the models to quickly obtain preliminary automatic diagnostic performance. The performance of the model was closely related to the data distributions of the training set. When the typical SRCs were few in the training set, the models would inevitably be biased towards other categories with more patients to obtain high accuracy in the evaluation \citep{AF2, AF8}. Therefore, when SRCs were mixed in the data of multiple subtypes of the gastrointestinal tract, special attention was needed to be paid to the problem of class imbalance, thereby improving the sensitivity of SRCs. However, fully supervised learning strongly relied on the fine-grained annotations, which required the pathologists to specify the location of each SRC. The time-consuming and laborious labeling requirements led to the scarcity of labeled data, which limited the scale of training data for fully supervised learning.

SRCs were difficult to be completely outlined even by experienced pathologists, so completely accurate fine annotations were almost impossible within limited time and cost. To ensure the accuracy of the annotations, some typical SRCs were carefully selected to teach the model learning and the confusing ones were ignored. Incomplete labeling was a ubiquitous phenomenon in SRC detection tasks \citep{DigestPath}. In addition to fully supervised learning with only limited annotations, semi-supervised learning further considered unlabeled SRCs \citep{AD2}. Incomplete labeling ensured the purity of the positive sample set, but could not guarantee the purity of the negative sample set, so it introduced a lot of noise to the learning of negative samples. Semi-supervised learning is an effective approach to reduce annotation pressure in the future, but the following two open issues need to be solved. First, how to obtain the high-dimensional representation of typical SRC features with only a few samples? Second, how to extract unlabeled true positive samples and eliminate false positive samples?

Weakly supervised learning can alleviate the dilemma of the scarcity of effective training data for fully supervised learning. Patient-level labels can be obtained directly from the current diagnosis pipeline. When a patient is diagnosed with SRCs, the corresponding screening images must contain SRCs although the specific location of each SRC is not specified. Weakly supervised learning embeds data with patient-level labels or image-level labels into the training set, greatly reducing the labeling pressure of pathologists and increasing the size of the training set. Combining fully and weakly supervised learning can balance the contradiction between SRC labeling pressure and training set size \citep{AL1, AW1}. In the future, weakly supervised algorithms should be mined deeply to make use of the SRC data containing patient-level labels already stored in various medical centers. In addition, additional SRC data collected over time may also be utilized to improve model performance without additional annotations.

Unsupervised learning is a valuable but also challenging training strategy for future research. A large amount of data without manual annotations can be utilized by unsupervised learning to actively cluster similar samples. In the early stage, the self-supervised learning strategy can train encoders through pretext tasks such as out-of-order correction and cloze \citep{MoCo}. The encoders with pretrained parameters are then fine-tuned on specific tasks, obtaining encoders that can extract SRC generalization features and well-performing decoders. In the later stage, ideal unsupervised learning will adaptively aggregate homogeneous tissues in the images into fine-grained subcategories, such as SRC nuclei, intracellular proteins, normal nuclei, and lymphocytes. Then, they are combined into target categories according to requirements, such as SRCs and lesion regions containing SRCs. 
\subsection{From algorithms to assistance}
  The algorithms designed for the DigestPath dataset detected typical SRCs in gastrointestinal histopathological patches \citep{DigestPath,AD2}, but were not suitable for clinically complex screening of SRCs in multiple organs. Most algorithms only accomplished SRC diagnosis for specific datasets, which ignored the attribute information such as the originated organs and invasion degrees, thus hindering subsequent prognosis. To complete the transition from algorithms to diagnostic assistance, future studies need to make contributions in the following four aspects.
  \begin{itemize}
  	\item Following the pathologists’ pipeline of diagnosing lesions from different fields of view by changing the magnification of the microscope, the intelligent algorithms may consider the multi-scale features. Kosaraju et al. distributed the detection of gastric lesions at 20$\times$ and 5$\times$ magnifications through two different branches \citep{AF2}. To simulate the visual diagnostic habits of the human eyes, more scales should be considered in future algorithms. Clustered SRCs can be detected at a faster speed at low magnification, while isolated SRCs will not be missed at high magnification.
  	\item Special attention could be provided to SRCs of metastases. Although SRCs occur mostly in the gastrointestinal tract, they may appear in various organs such as lung, pancreas, appendix, gallbladder, breast, and bladder \citep{S11}. For example, metastatic SRCs may stimulate tissue proliferation in the ovary, making SRCs more difficult to identify. When SRCs are implanted in the body cavity, omentum peritoneum, chest wall or abdominal wall of the patients, the surgical procedure and systemic treatment strategy will be affected. 
    \item \Rone{Tumor microenvironment of SRC carcinoma will be given attention. Current diagnostic models predominantly focus on SRC detection in isolation, neglecting the critical tumor microenvironment. Future algorithms should incorporate spatial co-localization analysis of peritumoral constituents, particularly tumor-infiltrating lymphocytes, macrophages, and stromal fibroblasts, to validate SRC identification through biologically plausible context. Quantifying spatial relationships would provide diagnostic corroboration while enabling microenvironment-driven prognostic stratification.}
  	\item The prognosis of SRCs will be directly correlated by algorithm-based lesion quantification. Prognosis through high-dimensional representation of SRCs in CT and MRI has been initially practiced and the feasibility was demonstrated \citep{AA1, AA3}. Since histopathology plays an important role in the diagnosis of SRCs, there is great research values and improvement room to realize prognosis prediction based on intelligent features. In addition, prognostic factors such as sampling location and depth of invasion may also be embedded in the intelligent analysis process in the future.
  	\item Future algorithms should be oriented to assistant needs throughout the entire diagnosis process. The current algorithms usually only solved a single problem in a simple scenario, such as SRC identification in gastric pathological slides. In the future, an end-to-end AI framework that fits the clinical diagnosis process will be welcomed, thereby promoting the implementation from algorithms to assistance.
  \end{itemize}
\subsection{Multi-modal diagnosis}
This survey exclusively addresses algorithms that leveraged machine learning or deep learning to automatically extract features from medical images, omics, and text. It omits the methods solely inferring prognosis directly from clinical information such as age, gender, and tumor characteristics, as these approaches still depended on manual interpretation \citep{B2336, B2337, B2339, B2342, B2354, B2367, B2374}. Meanwhile, methods involving manual delineation of RoIs in CT images for subsequent local feature extraction and fusion with clinical information for prognosis analysis were also excluded \citep{B2338, B2340, B2355}. Notably, both image-based and clinical information-based algorithms started from a single modality, disregarding the diagnostic potential of other modalities, which does not align with actual clinical diagnostic workflows. Therefore, AI algorithms should balance the multi-modal predicted results to improve the diagnostic accuracy. The great potential of multi-modal learning for cancer prognosis analysis has been demonstrated by a study using multi-modal learning to integrate and analyze WSIs and genetic maps of 14 cancers \citep{S15}. \Rone{In this survey, multi-modal studies of the collaboration between images and text} \citep{M2-ALPaCA,M2-CHIEF,M2-ConcepPath,M2-CONCH,M2-‌MR-PLIP,M2-MUSK,M2-OmniPath,M2-PRISM,M2-Prov-GigaPath,M2-TITAN} \Rone{collaboration between images and omics} \citep{M1-MISO,M1-OmiCLIP}, \Rone{collaboration of images, text, and omics} \citep{M3-GECKO,M3-ModalTune,M3-PCA,M3-spEMO}, \Rone{and other collaborations} \citep{MM-GAN,MM-Gemini} \Rone{have been involved and reviewed. Current multi-modal frameworks, whether pan-cancer or SRC-specific, largely conform to conventional AI training paradigms, inadequately embedding the distinctive pathobiology of SRC carcinoma. This limitation is clinically consequential: many gastric SRCs evade endoscopic biopsy detection due to their submucosal infiltration pattern, leading to false-negative diagnoses. To mitigate such underdiagnosis, future models must establish high-dimensional cross-modal mappings integrating patient-specific clinical profiles, laboratory biomarkers, radiological signature, sequential histopathological data from endoscopic biopsies, intraoperative frozen sections, and permanent specimens. Critically, the future multi-modal models should incorporate graceful degradation mechanisms, leveraging probabilistic graphical models or transformer-based attention gates, to maintain diagnostic robustness when modalities are missing or disturbed. }

\section{Conclusions}
\label{sec:conclusions}
This survey has systematically examined the advancements in AI-based SRC diagnosis from 2008 to \Rone{June 2025}, addressing critical gaps in the literature by integrating biological, technical, and clinical perspectives. \Rone{We categorized existing methodologies into unimodal (image-based classification, detection, segmentation, omics-based, and text-based analysis) and multi-modal approaches, highlighting their respective strengths and limitations in addressing the unique challenges of SRC identification, such as morphological diversity, image quality variability, and annotation incompleteness.} Key datasets like DigestPath and TCGA were analyzed, demonstrating both their utility and the need for larger, multi-institutional cohorts to improve model generalizability.

Despite significant progress, unresolved challenges remain, including the scarcity of high-quality annotated data, the integration of multi-scale histopathological features, and the translation of computational tools into routine clinical workflows. \Rone{Future research must prioritize the development of robust, interpretable AI systems that align with pathologists' diagnostic pipelines, incorporate tumor microenvironment analysis, and leverage multi-modal data fusion to bridge the gap between research and real-world implementation. The insights presented herein aim to guide future investigations, particularly for interdisciplinary teams seeking to advance intelligent diagnostic systems for this histopathologically complex and clinically aggressive carcinoma.}

\section*{CRediT authorship contribution statement}
\textbf{Zhu Meng}: Conceptualization, Investigation, Formal analysis, Visualization, Funding acquisition, Writing - original draft, Writing - review \& editing.
\textbf{Junhao Dong}: Conceptualization, Investigation, Formal analysis, Visualization, Writing - original draft, Writing - review \& editing.
\textbf{Limei Guo}: Conceptualization, Investigation, Formal analysis, Supervision, Writing - review \& editing.
\textbf{Fei Su}: Conceptualization, Supervision, Writing - review \& editing.
\textbf{Jiaxuan Liu}: Investigation, Visualization, Writing - review \& editing.
\textbf{Guangxi Wang}: Conceptualization, Investigation, Funding acquisition, Writing - review \& editing.
\textbf{Zhicheng Zhao}: Conceptualization, Supervision, Funding acquisition, Writing - review \& editing.

\section*{Declaration of competing interest}
The authors declare that they have no known competing financial interests or personal relationships that could have appeared to influence the work reported in this paper.

\section*{Acknowledgments}

This work was supported by Project funded by Chinese National Natural Science Foundation [grant numbers 62401069, U1931202, 62076033, 30700349, and 82102692], China Postdoctoral Science Foundation [grant number \\2023M730338], and The National Key Research and Development Program of China [grant number 2022YFC0868500].



\bibliographystyle{elsarticle-harv} 
\bibliography{ref}





\end{document}